\documentclass[useAMS,usenatbib]{mnras}

\usepackage{graphicx}
\usepackage{rotating}
\usepackage{amsmath}
\usepackage{amssymb}
\usepackage{lscape}
\usepackage{natbib}
\usepackage{keyval}
\usepackage{lastpage}
\usepackage{array}
\usepackage{dcolumn}
\usepackage{comment}
\usepackage{pifont}
\usepackage{natbib,times}
\usepackage[T1]{fontenc}
\usepackage{aecompl}

\newcommand{\mstar}{\hbox{${\mathrm M}_\ast$}}
\newcommand{\msun}{\hbox{$\rm{M_{_\odot}}$}}
\newcommand{\zsun}{\hbox{$\rm{Z_{_\odot}}$}}

\def\arcsec{\hbox{$^{\prime\prime}$}}

\def \be {\begin{equation}}
\def \ee {\end{equation}}

\def\gsim{\mathrel{\lower0.6ex\hbox{$\buildrel {\textstyle >}
 \over {\scriptstyle \sim}$}}}
\def\lsim{\mathrel{\lower0.6ex\hbox{$\buildrel {\textstyle <}
 \over {\scriptstyle \sim}$}}}

\def \Vmax {$\mathrm{V_{max}}$\,}

\title[The mechanisms for quiescent galaxy formation]{Galaxy And Mass Assembly (GAMA): The mechanisms for quiescent galaxy formation at $z<1$}

\author[K. Rowlands et al.]
{K. Rowlands$^{1,2}$\thanks{E-mail:katerowlands.astro@gmail.com}, V. Wild$^1$, N. Bourne$^3$, M. Bremer$^4$, S. Brough$^5$, S. P. Driver$^{6,1}$,
\newauthor A. M. Hopkins$^5$, M. S. Owers$^{7,5}$, S. Phillipps$^4$, K. Pimbblet$^{8,9}$, A. E. Sansom$^{10}$,
\newauthor L. Wang$^{11,12}$,
M. Alpaslan$^{13}$, J. Bland-Hawthorn$^{14}$, M. Colless$^{15}$, B. W. Holwerda$^{16,17}$,
\newauthor E. N. Taylor$^{18}$ \\
$^1$(SUPA) School of Physics \& Astronomy, University of St Andrews, North Haugh, St Andrews, Fife, KY16 9SS, UK \\
$^2$Department of Physics \& Astronomy, Johns Hopkins University, Bloomberg Center, 3400 N. Charles St., Baltimore, MD 21218, USA \\
$^3$Institute for Astronomy, University of Edinburgh, Royal Observatory, Blackford Hill, Edinburgh EH9 3HJ, UK \\
$^4$Astrophysics Group, School of Physics, University of Bristol, Tyndall Avenue, Bristol BS8 1TL, UK \\
$^5$Australian Astronomical Observatory, PO Box 915, North Ryde, NSW 1670, Australia \\
$^6$ICRAR, The University of Western Australia, 35 Stirling Highway, Crawley, WA 6009, Australia \\
$^7$Department of Physics and Astronomy, Macquarie University, NSW 2109, Australia \\
$^8$E. A. Milne Centre for Astrophysics, University of Hull, Cottingham Road, Kingston-upon-Hull, HU6 7RX, UK \\
$^9$School of Physics and Astronomy, Monash University, Clayton, Victoria 3800, Australia \\
$^{10}$Jeremiah Horrocks Institute, University of Central Lancashire, PR1 2HE Preston, UK \\
$^{11}$SRON Netherlands Institute for Space Research, Landleven 12, NL-9747 AD Groningen, the Netherlands \\
$^{12}$Kapteyn Astronomical Institute, University of Groningen, Postbus 800, NL-9700 AV Groningen, the Netherlands \\
$^{13}$NASA Ames Research Center, N232, Moffett Field, Mountain View, CA 94035, United States \\
$^{14}$Sydney Institute for Astronomy, School of Physics A28, University of Sydney, NSW 2006, Australia \\
$^{15}$Research School of Astronomy and Astrophysics, Australian National University, Canberra, ACT 2611, Australia \\
$^{16}$University of Leiden, Sterrenwacht Leiden, Niels Bohrweg 2, NL-2333 CA Leiden, The Netherlands \\
$^{17}$Department of Physics and Astronomy 102 Natural Science Building, University of Louisville, Louisville KY 40292, USA \\
$^{18}$Centre for Astrophysics and Supercomputing, Swinburne University of Technology, Hawthorn 3122, Australia
}

\begin{document}
\maketitle
\begin{abstract}
One key problem in astrophysics is understanding how and why galaxies switch off their star formation, building the quiescent population that we observe in the local Universe. From the GAMA and VIPERS surveys, we use spectroscopic indices to select quiescent and candidate transition galaxies. We identify potentially rapidly transitioning post-starburst galaxies, and slower transitioning green-valley galaxies. Over the last 8\,Gyrs the quiescent population has grown  more slowly in number density at high masses ($\mstar>10^{11}\msun$) than at intermediate masses ($\mstar>10^{10.6}\msun$). There is evolution in both the post-starburst and green valley stellar mass functions, consistent with higher mass galaxies quenching at earlier cosmic times. At intermediate masses ($\mstar>10^{10.6}\msun$) we find a green valley transition time-scale of 2.6\,Gyr. Alternatively, at $z\sim0.7$ the entire growth rate could be explained by fast-quenching post-starburst galaxies, with a visibility time-scale of 0.5\,Gyr. At lower redshift, the number density of post-starbursts is so low that an unphysically short visibility window would be required for them to contribute significantly to the quiescent population growth. The importance of the fast-quenching route may rapidly diminish at $z<1$. However, at high masses ($\mstar>10^{11}\msun$), there is tension between the large number of candidate transition galaxies compared to the slow growth of the quiescent population. This could be resolved if not all high mass post-starburst and green-valley galaxies are transitioning from star-forming to quiescent, for example if they rejuvenate out of the quiescent population following the accretion of gas and triggering of star formation, or if they fail to completely quench their star formation.
\end{abstract}

\section{Introduction}
The galaxy population displays a colour and morphological bimodality
\citep{Strateva01, Blanton03, Baldry04, Bell04b}, which emerged at $z<2$ \citep[e.g.][]{Arnouts07, Brammer11, Wuyts11, Mortlock13, Whitaker15}. Wide-area galaxy surveys have shown that the stellar mass density of the star forming population has been approximately constant over the last 8\,Gyr ($z<1$, \citealp[e.g.][]{Pozzetti10, Ilbert13, Moustakas13, Muzzin13}). These recent studies have also charted the growth of the quiescent population over cosmic time, although discrepancies exist at $z<1$ as to how quickly the quiescent population grows. Many studies found that the quiescent population doubled in mass between $0<z<1$ \citep{Bell04, Brown07, Arnouts07}. Integrating over galaxies of all masses, \citet{Muzzin13} found that the quiescent population grows in stellar mass density from $z=1$ to $z=0.3$, but using the same survey data, \citet{Ilbert13} found that the number density of quiescent galaxies is flat from $z=1$ to the present day. The growth rate of the quiescent population is likely to be mass dependent; \citet{Moustakas13} concluded that the number density of quiescent galaxies grows from $z=1$ to now for low mass ($\mstar<10^{10.6}\msun$), but not for higher mass galaxies. For the quiescent population to grow, galaxies must transform from star-forming to quiescent, as quiescent galaxies are no longer forming stars. Understanding the processes which quench star formation, and the time-scale over which this happens, is one of the major open questions in extragalactic astronomy.

There is much debate about the dominant quenching mechanisms and transition time-scales for galaxies. There are two main quenching channels suggested to halt star-formation in galaxies: fast and slow, and while there is no agreement on exactly \emph{how fast} or \emph{how slow} these channels are, they are generally linked to different quenching processes \citep{Faber07, Fang12, Fang13, Barro13, Yesuf14}. Star formation in galaxies could quench slowly over many Gyr, where the gas may be stabilised against collapse (e.g. morphological quenching, \citealp{Martig09}), or the supply is cut off and galaxies gradually exhaust their gas through star formation over a time-scale of a few Gyr. For galaxies to stop forming stars more rapidly requires the removal of large amounts of gas. Mergers could be responsible for triggering a chain of events which lead to a more rapid shutdown of star-formation in galaxies. Models have shown that the torques induced during a gas-rich major merger might funnel gas towards the galaxy centre, triggering an intense burst of star formation \citep[e.g.][]{Mihos_Hernquist94, Mihos_Hernquist96, Barnes_Hernquist96}, capable of consuming a significant portion of a galaxy's gas supply. The gas is then rapidly depleted, and may additionally be prevented from forming stars via feedback mechanisms \citep[e.g.][]{Benson03, DiMatteo05} from stellar or AGN-driven winds \citep[e.g.][]{Springel05, Hopkins07, Khalatyan08, Kaviraj11}. Other environment-dependant mechanisms such as ram pressure stripping \citep{Gunn_Gott1972, McCarthy08} may also remove the gas reservoir on short-intermediate time-scales.

Observational results on quenching time-scales and mechanisms vary substantially. The dearth of galaxies in the region intermediate between the star-forming and quiescent populations in the optical/UV colour-magnitude diagram has often been used to argue that galaxies transition rapidly from star-forming to quiescent \citep[e.g.][]{Martin07, Kaviraj07}. Using broad-band colours, \citet{Schawinski14} concluded that disc galaxies quench slowly over many Gyrs via gentle, secular processes with little morphological change, whereas spheroidal galaxies undergo faster, more violent quenching which also transforms their morphology. By fitting chemical evolution models to the difference in stellar metallicity between star forming and quiescent galaxies, \citet*{Peng15} found that $\mstar<10^{11}\msun$ galaxies in the local Universe are quenched over a time-scale of 4\,Gyr, which suggests strangulation is the dominant mechanism, whereby halo gas is removed as a galaxy falls into a group/cluster. \citet{Wetzel13} found that satellite galaxies continue to form stars for 2--4\,Gyr before quenching rapidly in $<0.8$\,Gyr, again leading them to suggest that gas exhaustion (i.e. strangulation) of the gas reservoir is the primary quenching mechanism. \citet{Haines13} concluded that cluster galaxies are quenched upon infall on time-scales of 0.7--2.0\,Gyr, and that slow quenching is suggestive of ram-pressure stripping or starvation mechanisms. The observed decrease in the fraction of star-forming galaxies with increasing environmental density and the independence of SFR and environment \citep{Wijesinghe12, Robotham13} suggests that galaxy transformation due to environmental processes must be rapid or have happened long ago \citep{Brough13}. Cosmological simulations are also starting to provide constraints: \citet{Trayford16} found in the Evolution and Assembly of GaLaxies and their Environments (EAGLE) simulation that the majority of green valley galaxies transition over a $<2$\,Gyr time-scale. In reality, there is likely to be a diversity in quenching time-scales for galaxies even in the local Universe \citep{Smethurst15}, see also \citet*{McGee14} for a compilation of quenching time-scale estimates.

It is clear that the relative importance of the fast and slow quenching channels are not well known, and may change over cosmic time, with stellar mass, and environment \citep{Peng10, Wijesinghe12, Crossett16, Hahn16}. Such variation may help to explain the diversity of observational results, however, observational methods for identifying quenched and transition galaxies may also be partly responsible. Previous studies have largely relied on broad-band photometric data, with any available spectroscopic data only used to provide a redshift to help with the correction of observed frame colours and environment estimates. Good quality spectroscopic data of galaxy continua contain a wealth of information on the star formation history of galaxies, and are arguably better suited to cleanly identifying both fully quenched and transitioning galaxies. In this paper we fully exploit the spectroscopic data from the Galaxy And Mass Assembly (GAMA) survey and VIMOS Public Extragalactic Redshift Survey (VIPERS) to robustly identify fully quenched and candidate fast and slow quenching galaxies.

To study galaxies undergoing fast quenching we need galaxies where we have a good constraint on their recent star-formation history. Post-starburst galaxies, where a galaxy has recently undergone a starburst followed by quenching in the last 1\,Gyr, are ideal for studying fast quenching. Post-starburst galaxies (PSBs) are sufficiently common at $z\sim1-2$ that they may contribute significantly to the growth of the red-sequence at this important epoch \citep{Wild16}. It is not well known how much PSBs contribute to the build-up of the quiescent population at $z<1$, due to small number statistics in previous redshift surveys \citep{Blake04, Wild09, Vergani10}, and aperture bias in spectroscopic surveys at very low redshifts \citep{Brough13, Iglesias_Paramo13, Richards16}. Furthermore, studies of the evolution of the quiescent and green valley populations have commonly been done using broad-band photometry. In such studies the sample selection and physical properties can be affected by dust, and there is a larger uncertainty on parameters such as stellar population age, stellar mass and photometric redshift compared to spectroscopic studies. Using spectra allows us to cleanly classify galaxies according to their likely quenching time-scales. By identifying large numbers of PSB and green valley galaxies in large spectroscopic surveys, we can identify which quenching channels are important for building the quiescent population at low redshift.

In this paper we investigate the mass functions and number density evolution of candidate transition and quenched galaxies at $0<z<1$. This allows us to investigate whether the quiescent galaxy population is growing at $z<1$, and what galaxies are responsible for any growth. We adopt a cosmology with $\Omega_m=0.30,\,\Omega_{\Lambda}=0.70$ and $H_o=70\,
\rm{km\,s^{-1}\,Mpc^{-1}}$.

\section{Data}
Due to the rarity of PSBs in the local Universe, large-area spectroscopic surveys are required to identify them. Our study necessitates spectra so we can robustly identify quiescent and transition galaxies, a high spectroscopic completeness and a good understanding of the survey selection function. These requirements are met by the GAMA and VIPERS surveys. The GAMA survey allows us to span the range $0.05<z<0.35$, above which only the most massive galaxies have adequate SNR spectra. The VIPERS data allows us to extend our study to higher redshift from $0.5<z<1$. Together these surveys give a total timespan of 6.5\,Gyr ($0.05<z<1.0$) to study galaxy evolution.

\subsection{GAMA}
The GAMA survey \citep{Driver11, Liske15} is a multiwavelength photometric and redshift database, covering 230\,deg$^2$ in three equatorial fields at $\sim9$, 12 and 14.5 hours (G09, G12 and G15), and two southern regions (G02 and G23). The GAMA database provides $r$-band defined matched aperture photometry from the UV--far-infrared as described in \citet{Hill11, Driver16} and \citet{Wright16}. In this work we use the equatorial regions as they are the most spectroscopically complete to $r=19.8$ mag, which cover 180\,deg$^2$.

Spectra are obtained for $\sim250,000$ galaxies with a magnitude limit of $r_{AB}<19.8$ mag mostly using the AAOmega spectrograph \citep{Saunders04, Sharp06} at the Anglo Australian Telescope. The AAOmega spectra \citep{Hopkins13} have a wavelength range of $3750-8850$\AA\ and a resolution of $R\sim1100$ at $\sim4000$\AA. Additional spectra are included from the Sloan Digital Sky Survey \citep[SDSS,][]{SDSS_York2000}, which have a wavelength range of $3700-9200$\AA\ and a resolution of $R\sim1600$ at $\sim4000$\AA. The physical scale covered by the 2\arcsec{} AAOmega fibres range from 2.0\,kpc at $z=0.05$ to 9.9\,kpc at $z=0.35$. The 3\arcsec{} SDSS fibres cover 2.9\,kpc at $z=0.05$ and 14.8\,kpc at $z=0.35$. We discuss the effects of aperture bias in Section~\ref{sec:Aperture_bias}, but we note that it is minimised by excluding galaxies at $z<0.05$. We do not include GAMA spectra from surveys such as 6dFGS which are not flux calibrated \citep{Hopkins13}.

We include all GAMA II main survey galaxies which have science quality redshifts ($nQ>2$), $10.0<r_{\mathrm{Petro}}<19.8$ mag and $9.9<\mathrm{log}_{10}(\mstar)<12$ totalling 111477 spectra from $0.05<z<0.35$. These include 97872 GAMA spectra and 13605 SDSS spectra. From this sample we then excluded 1761 problematic spectra which show, e.g. fibre fringing (identified by eye and through GAMA redshift catalogue flags) and 331 galaxies hosting broad-line AGN from \citet{Gordon17} and \citet{Schneider07}, which prevents us from robustly measuring spectral features.

In this work we calculate stellar masses using photometry from the GAMA {\sc lambdar} Data Release \citep{Driver16, Wright17}. The catalogue comprises deblended matched aperture photometry in 21 bands from the  observed frame $FUV$-FIR, with measurements accounting for differences in pixel scale and PSF in each band. We utilise the UV $FUV$ and $NUV$ \emph{GALEX} data, optical $ugri$ magnitudes from SDSS DR6 imaging \citep{AM09} and near-infrared $ZYJHK$ photometry from the Visible and Infrared Telescope for Astronomy (VISTA, \citealp{Sutherland15}), as part of the VIsta Kilo-degree INfrared Galaxy survey (VIKING). All photometry has been galactic extinction corrected using the values of $E(B-V)$ derived using the \citet{Schlegel98} Galactic extinction maps for a total-to-selective extinction ratio of $R_V = 3.1$. For all fluxes we convolve the catalogue error in quadrature with a calibration error of 10\% of the flux respectively, to allow for differences in the methods used to measure total photometry and errors in the spectral synthesis models used to fit the underlying stellar populations.

The GAMA and SDSS spectra were taken at a much higher spectral resolution ($R\sim1100$, and $R\sim1600$, respectively) than the VIPERS spectra ($R\sim210$). To perform a consistent analysis, we convolve the GAMA/SDSS spectra to the same spectral resolution as the VIPERS spectra using a Gaussian convolution kernel. During the convolution we linearly interpolate over bad pixels. In practise this makes little difference to the galaxy spectra, but does allow us to include more spectra in our analysis which would have important spectral features masked out if we simply propagated the bad pixels in the convolution. We measure the new errors for each convolved spectrum by scaling the unconvolved error array to the standard deviation of the flux in line-free regions of the convolved spectrum at $4200-4300$\AA, to account for covariance between smoothed spectral pixels.

From the GAMA sample we select galaxies to be at $z>0.05$ so that the higher-order Balmer lines (H$\delta$, H$\epsilon$, etc.) are redshifted into a more sensitive portion of the AAOmega spectrograph, and away from regions at shorter wavelengths where poor flat fielding can affect the spectra. Additionally, at $z>0.05$ the fibre samples a substantial fraction of the galaxy light (10--30\% of the Petrosian radius) and so minimises aperture effects (see Section~\ref{sec:Aperture_bias} and \citealt{Kewley05}). The upper redshift limit is set to $z=0.35$ as above this the mass completeness limit exceeds $\mstar>10^{11}$ in some spectral classes, leaving us with few galaxies to study. Note that the 3750--4150\AA\ region (used for spectral classification, see Section~\ref{sec:sample}) is required to always be in the observed spectral range.

\subsection{VIPERS}
The VIPERS Public Data Release 1\footnote{http://vipers.inaf.it/} provides 61221 spectra for galaxies with $17.5<i_{AB}<22.5$ mag.
The PDR1 covers 10.315\,deg$^2$ (after accounting for the photometric and spectroscopic masks) in the Canada-France-Hawaii Telescope Legacy Survey Wide (CFHTLS-Wide) W1 and W4 fields. A colour selection using $(g-r)$ and $(r-i)$ cuts was used to primarily select galaxies in the range $0.5<z<1.3$. Spectra were observed using the VIMOS spectrograph on the VLT with the LR-grism, yielding a spectral resolution of $R\sim210$ (at $\sim6000$\AA) with wavelength coverage from $5500-9500$\AA.
Further details of the survey data are given in \citet{Guzzo14} and \citet{Garilli14}. The VIPERS survey used a slit with 1\arcsec{} width, but with considerably longer length. The majority of each galaxy should be in each slit and any aperture bias between the GAMA and VIPERS samples should be negligible, except at the lowest redshifts in the GAMA sample.

To calculate stellar masses (see Section~\ref{sec:stellar_masses}), we use total broad-band photometry in the $FUV$, $NUV$, $u\ast$, $g'$, $r'$, $i'$, $z'$ and $K_s$ bands measured using SExtractor {\sc MAG\_AUTO} as described in \citet{Moutard16}. All photometry has been galactic extinction corrected using $E(B-V)$ values of 0.025 in the W1 field and 0.05 in the W4 field \citep{Fritz14}, derived using the \citet{Schlegel98} Galactic extinction maps for a total-to-selective extinction ratio of $R_V = 3.1$. WIRCAM $K_s$ band data is available for 91.5\% galaxies. We checked that the lack of NIR data for some galaxies does not significantly affect our stellar mass estimates. For all fluxes we convolve the catalogue error in quadrature with a calibration error of 10\% of the flux respectively, to allow for differences in the methods used to measure total photometry and errors in the spectral synthesis models used to fit the underlying stellar populations.

We use galaxies with $0.5<z<1.0$, $9.9<\mathrm{log}_{10}(\mstar)<12$, and which have secure spectroscopic redshifts with flags $2.0 \leq z_\mathrm{flg} \leq 9.5$ (corresponding to a 95\% confidence limit on the redshift) and which are inside the photometric mask. 616 galaxies with broad-line AGN ($10<zflg<20$ or agnFlag $ = 1$) were excluded from our analysis. We restrict the upper redshift limit of the VIPERS survey to $z=1.0$, as above this redshift we are only mass complete to the most massive galaxies ($\mstar>10^{11.5}$) which are not the main subject of this study. Our final sample comprises 29734 galaxies on which to perform spectroscopic classification.

\subsection{Spectroscopic Sample Classification} \label{sec:sample}
In the integrated optical fibre spectrum of a galaxy the signatures of
stars of different ages can be used to obtain information about a galaxy's
recent star-formation history (SFH). To define our sample we make use of two
particular features of optical spectra: the 4000\AA\ break strength and Balmer absorption line strength. Following the method outlined in \citet{Wild07, Wild09}, we define two spectral indices which are based on a Principal Component Analysis (PCA) of the 3750--4150\AA\ region of the spectra. PC1 is the strength of the 4000\AA\ break (equivalent to the $D_{n}4000$ index), and PC2 is excess Balmer absorption (of all Balmer lines simultaneously) over that expected for the 4000\AA\ break strength. The eigenbasis that defines the principal components is taken from \citet{Wild09}, and was built using observed VVDS spectra.

To calculate the principal component amplitudes for each spectrum, we correct for Galactic extinction using the \citet{Cardelli89} extinction law, shift to rest-frame wavelengths and interpolate the spectra onto a common wavelength grid. We then project each spectrum onto the eigenbasis using the `gappy-PCA' procedure of \citet{Connolly_Szalay99} to account for possible gaps in the spectra. Pixels are weighted by their errors during the projection, and gaps in the spectra due to bad pixels are given zero weight. The normalisation of the spectra is also free to vary in the projection using the method introduced by \citet{Wild07}.

\begin{figure*}
\includegraphics[width=1.0\textwidth]{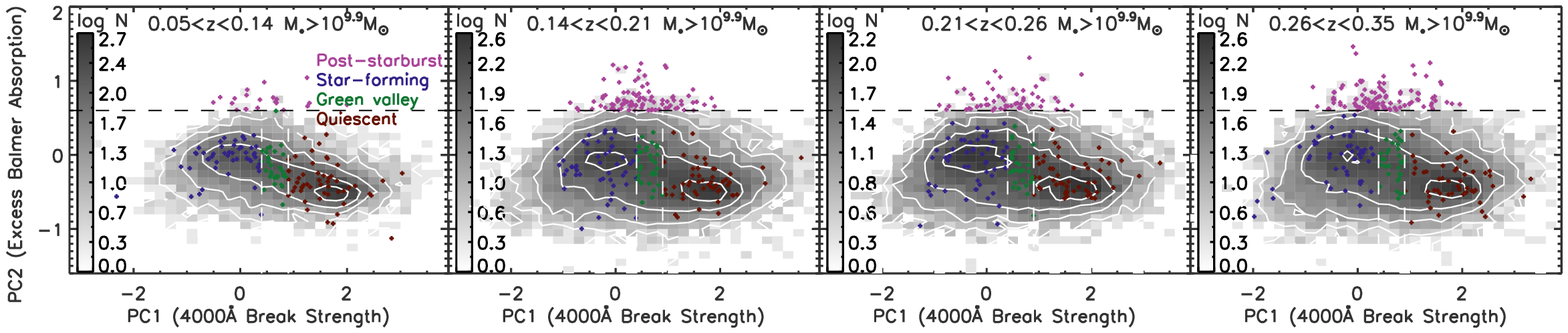}
\includegraphics[width=0.8\textwidth]{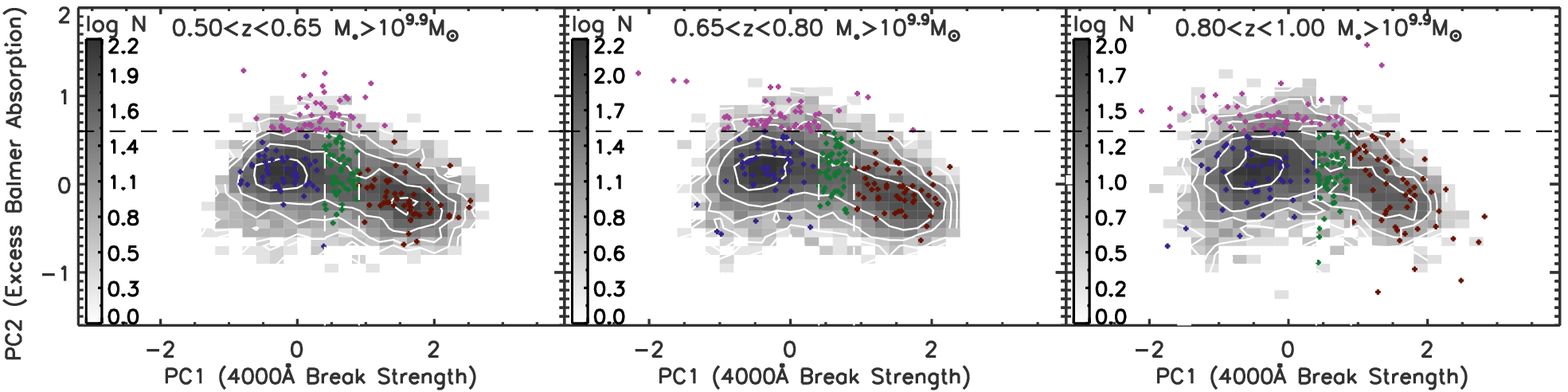}
\caption{The distribution of the 4000\AA\ break strength (PC1) and excess Balmer absorption (PC2) as measured by a principal component analysis of the 4000\AA\ spectral region of the GAMA and VIPERS galaxies, in the seven different redshift bins used in this work. The grey-scale indicates the logarithmic number of objects. The coloured dots are random samples of galaxies which occupy each spectral class delineated by dashed lines: quiescent (red), star forming (blue), green valley (green) and PSB (purple); these are discussed in detail in Section~\ref{sec:sample}. Contours show 10, 30, 50, 70 and 90\% of the maximum number of galaxies in the sample.}
\label{fig:GAMA_pc12}
\end{figure*}

In Figure~\ref{fig:GAMA_pc12} we show the distribution of the two spectral indices for galaxies in the GAMA and VIPERS surveys which have SNR per 6\AA\ pixel $>6.5$ at $\sim4000$\AA. This choice of SNR cut allows us to reliably measure spectral indices from low resolution spectra \citep{Wild09}. Our sample comprises 70668 and 21519 galaxies from GAMA and VIPERS, respectively.

In Figure~\ref{fig:GAMA_pc12} we divide our sample into four spectral classes based on their values of PC1 and PC2. The boundaries between the spectral classes are red: PC1$>0.9$, green: $<0.4<$PC1$<0.9$, star-forming: PC1$<0.4$ and PSB: PC2$>0.6$. Classification is not influenced in any way by commonly used star-formation indicators such as [OII] and H$\alpha$ fluxes.
After a starburst, the Balmer absorption lines increase in strength as the galaxy passes into the PSB phase \citep{DresslerGunn83, CouchSharples87} i.e. A/F star light dominates the integrated galaxy spectrum. These objects with stronger Balmer absorption lines compared to their expected 4000\AA\ break strength lie to the top of each panel in
Figure~\ref{fig:GAMA_pc12}. The boundaries for the PSB class are defined to select the population outliers with high PC2. At low redshift, there are very few PSBs in each of the four redshift bins in the GAMA survey, so we collapse all of the PSBs into one large redshift bin from $0.05<z<0.26$ so that we have sufficient number statistics for our analysis (see Figure~\ref{fig:GAMA_pc12_unclean}). We cannot extend the PSB sample to the highest redshift range of the GAMA sample as our mass completeness drops below our 90\% limit (see Section~\ref{sec:mass_completeness}) for \mstar$>10^{10.6}$\msun. We visually inspected all of the candidate PSB spectra above our SNR limit. As shown in Appendix~\ref{sec:Unclean_PC12}, we found that $\sim 2/3$ of GAMA galaxies with PC2$>0.6$ are contaminants caused by problems with unmasked noise spikes, or exhibited an extreme fall off in flux to the blue (this could be due to poor tracing of the fibre flux on the CCD when the SNR is low). Furthermore, some spectra in the PSB region were removed if we could not positively identify a Balmer series. We note that if we did not remove the visually identified contaminants from the PSB sample our conclusions would be unchanged, even if the PSBs are twice as numerous at low redshift.

One concern is that, as we exclude broad line AGN from our samples and PSBs are found to contain a higher fraction of narrow-line AGN than other galaxies \citep{Yan06,Wild07}, we may be systematically missing PSBs from our samples. However, a typical AGN lifetime is two orders of magnitude shorter than the time during which PSB features are visible \citep[e.g][]{Martini_Weinberg01}, thus even if all PSBs undergo a powerful unobscured AGN phase (which we consider unlikely at the redshifts studied in this paper) we will only miss a small fraction of galaxies.

Galaxies which show no evidence of recent or current star formation comprise the quiescent population which lies on the right in each panel of Figure~\ref{fig:GAMA_pc12}, as they have a strong 4000\AA\ break. Galaxies that are forming stars lie in the centre and left of each panel. These galaxies have younger mean stellar ages and therefore weaker 4000\AA\ breaks.
Galaxies in the sparsely populated region between the star-forming and quiescent populations are defined as green valley (akin to that of the green valley in NUV/optical colour magnitude diagrams). These spectroscopic green valley galaxies do not show characteristic deep Balmer absorption lines, which indicates a slower transition for these galaxies compared to PSB galaxies.
We make fixed cuts in PC1 and PC2 to separate our spectral classes, and we do not evolve these with redshift. This is because we want to select candidate transition populations between defined limits (i.e. fixed age) to test whether galaxies are changing from star forming to quiescent through these transition populations.

\begin{figure*}
\begin{center}
\begin{minipage}[t]{1.0\textwidth}\
\begin{center}
$\begin{array}{cc}
\includegraphics[width=0.45\textwidth]{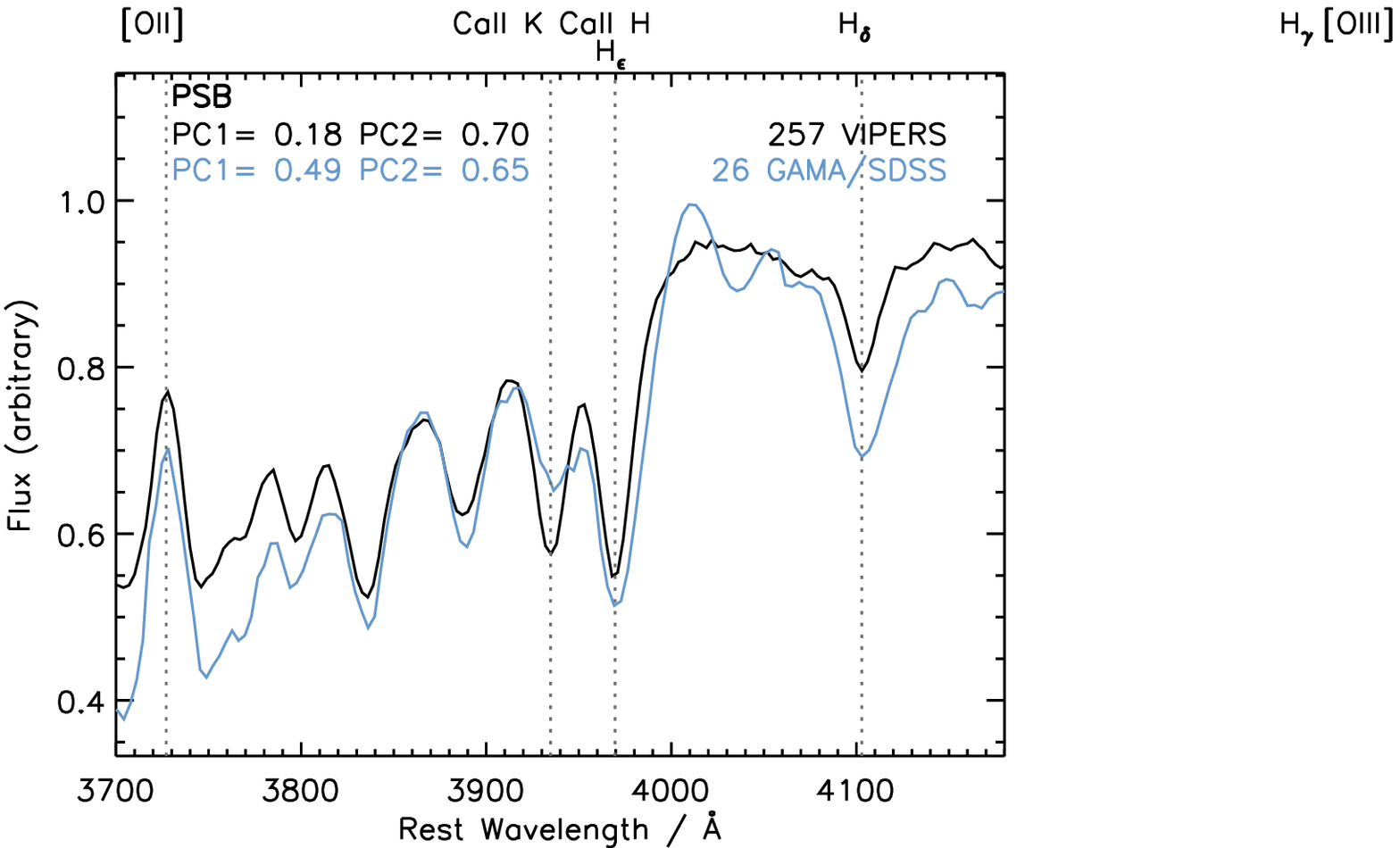} &
\includegraphics[width=0.45\textwidth, clip=true, trim=0mm 0mm 0mm 0mm]{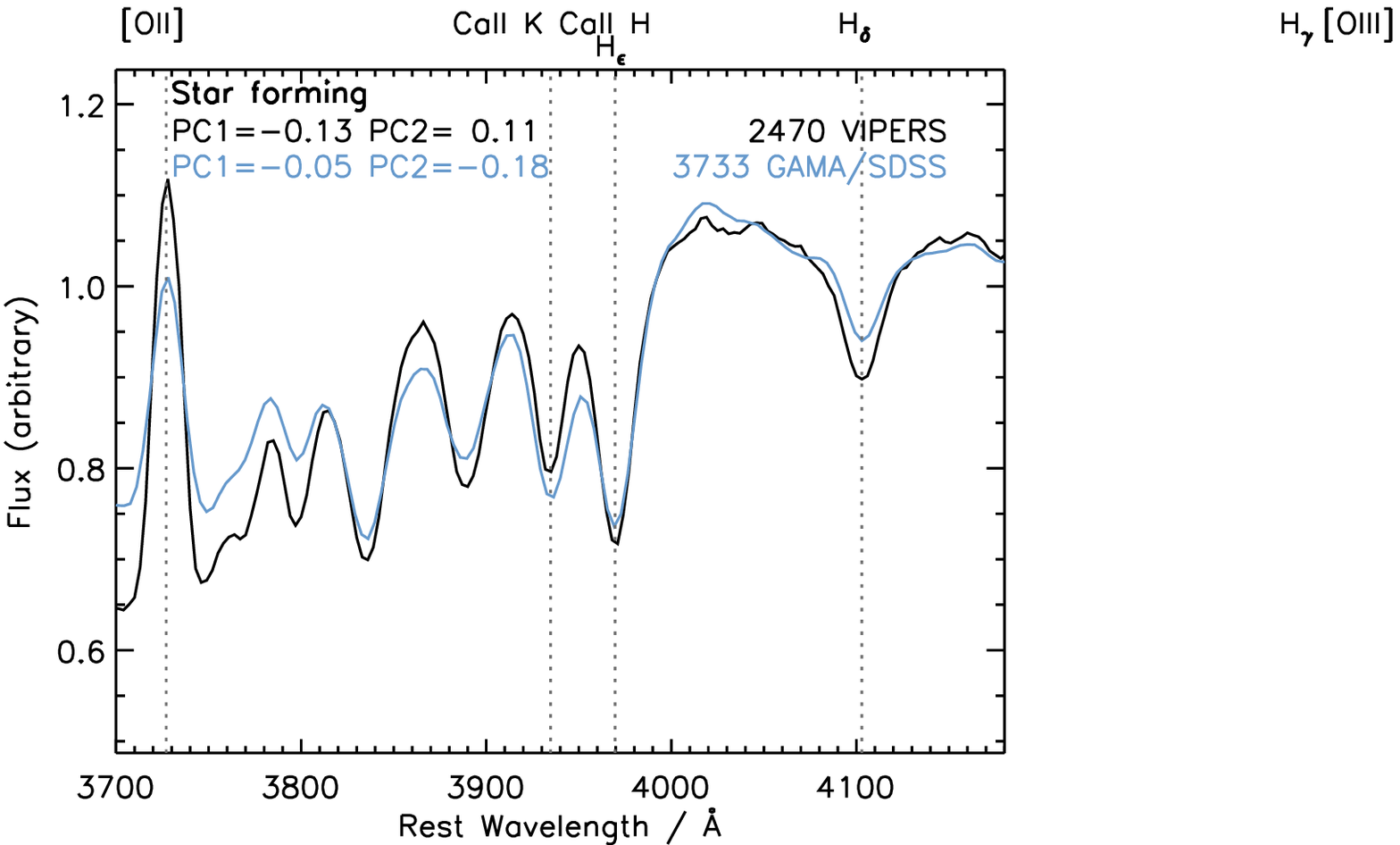} \\
\end{array}$
$\begin{array}{cc}
\includegraphics[width=0.45\textwidth]{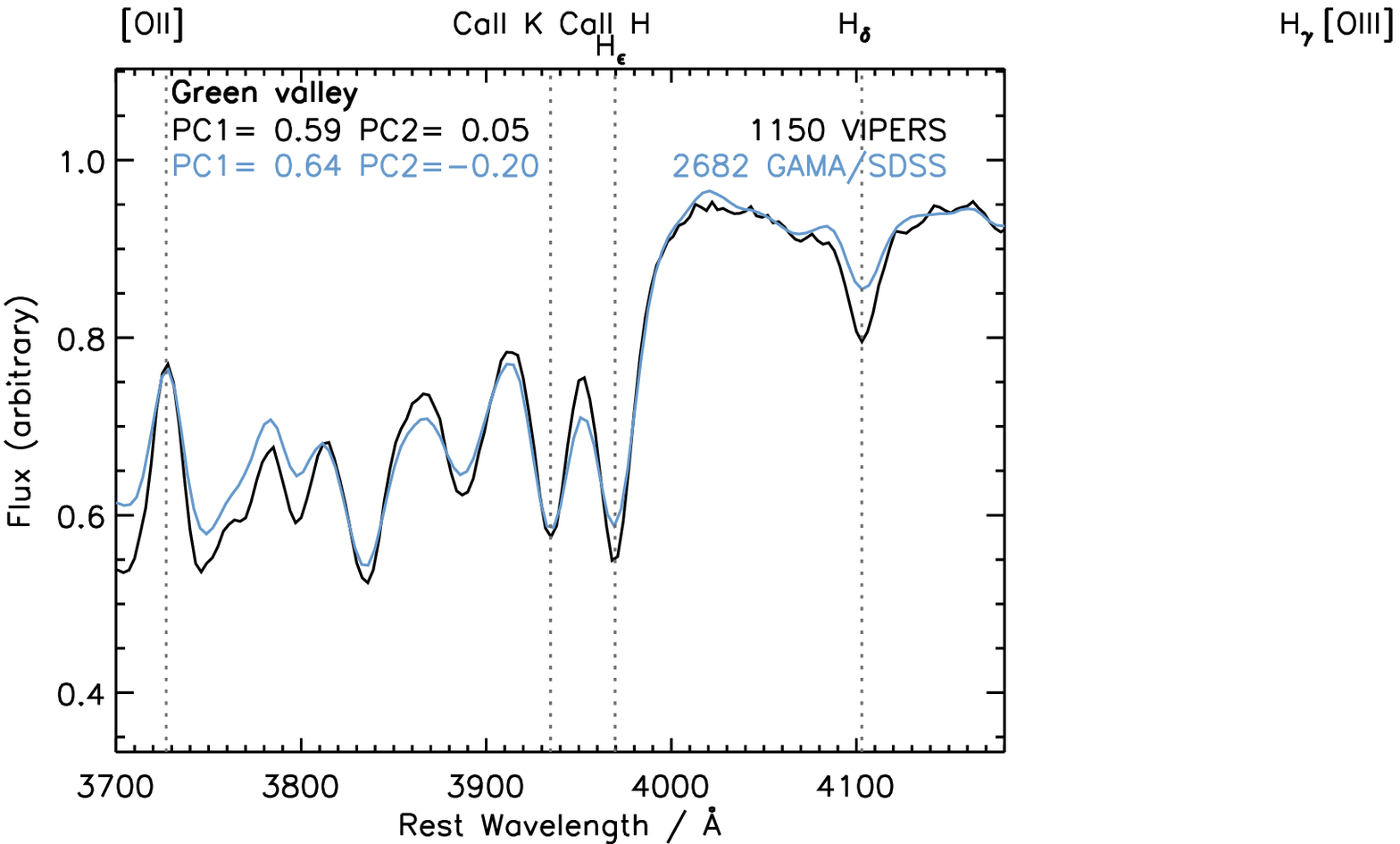} &
\includegraphics[width=0.45\textwidth, clip=true, trim=0mm 0mm 0mm 0mm]{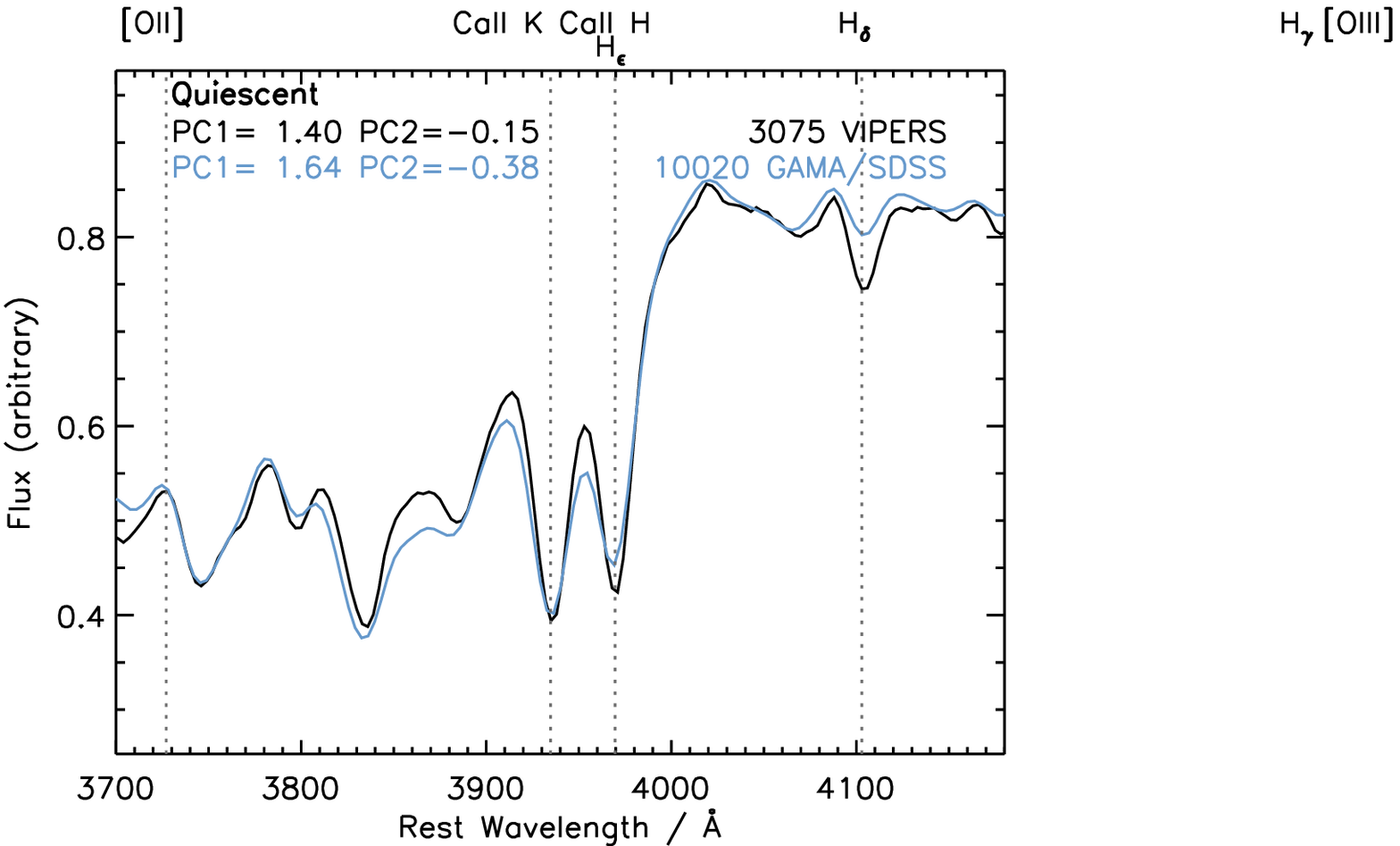}  \\
\end{array}$
\end{center}
\end{minipage}
\end{center}
\caption{Stacked spectra in each spectral class with $10.6<\mathrm{log}_{10}(\mstar)<11$. We stack galaxies in a fixed mass range so that we can be sure we are comparing similar galaxies in each survey. Black lines show the stacked VIPERS spectra with $0.5<z<0.6$, blue lines show the stacked GAMA/SDSS spectra with $0.05<z<0.35$ convolved to the same resolution as the VIPERS spectra. Note that the PSBs are selected to be at $0.05<z<0.26$. The stacked spectra are normalised to the same value at $4000$\AA\ to aid comparison of the two samples. Dotted vertical lines indicate the rest-frame vacuum wavelengths of emission and absorption lines labelled at the top of each panel.}
\label{fig:stacked_spectra}
\end{figure*}

The boundaries between the spectral classes are somewhat arbitrary, but the broad-band colours of spectroscopically selected galaxies lie in the expected regions of the $g-r$ colour-magnitude diagram (see Figures~\ref{fig:colour_mag} and \ref{fig:UVJ}). Stacking spectra in each class with similar stellar masses shows that on average the galaxies show the expected characteristic features, see Figure~\ref{fig:stacked_spectra}. Star-forming galaxies show strong emission lines, a weak 4000\AA\ break, and blue continua. The stacked quiescent galaxies show strong 4000\AA\ breaks and no emission lines. Green valley galaxies show spectra intermediate between those of star-forming and quiescent galaxies with moderately strong 4000\AA\ breaks and weak emission lines.
PSBs have strong Balmer absorption lines and moderately strong 4000\AA\ breaks. Our stacked PSB spectra show a strong [OII] emission line; we measure equivalent widths (EWs) of $-10.8$\AA\ and $-10.1$\AA\ for the stacked GAMA and VIPERS spectra, respectively. Note that our selection method makes no cuts on emission line strength, as is often done in the selection of PSBs \citep{Goto05, Goto08}. If we were to use the \citet{Goto08} cut of [OII] EW$>-2.5$\AA, the average PSB in our sample would be excluded. It is important not to exclude galaxies with emission lines, as narrow line AGN are common in PSB samples \citep{Wild07, Yan06, Yan09}, and shocks can excite emission lines in PSBs \citep{Alatalo16a}. We defer examination of the ionising sources in PSBs to a future paper.

The stacked spectra in each spectral class look very similar in the 4000\AA\ break region for both the GAMA/SDSS and VIPERS samples. The similarity is quantified by the values of PC1 and PC2 of the stacked spectra for the GAMA and VIPERS samples in Figure~\ref{fig:stacked_spectra}. This shows that our PCA method is successful at selecting similar galaxies in each sample, despite differences in redshift and the initial spectral resolution. We observe a slightly weaker 4000\AA\ break and stronger Balmer absorption lines in the stacked VIPERS spectra, indicting that on average the higher redshift galaxies are younger. This is most pronounced in the stacked star forming spectra, where a significantly stronger [OII] line is visible in the VIPERS spectra compared to GAMA spectra consistent with the expected increase in the specific SFR (SSFR) of galaxies with redshift.

\subsection{Stellar masses}
\label{sec:stellar_masses}
Stellar masses were calculated for each galaxy using
a Bayesian analysis which accounts for the degeneracy between
physical parameters. Specifically, we fit a library of tens of thousands (depending on the redshift) of \citet{BC03} population synthesis models to the
$FUV-K$ broadband photometry, to obtain a probability density function (PDF) for each physical property.
The model libraries have a wide range of star-formation histories,
two-component dust contents \citep{Charlot_Fall00} and metallicities from $0.5-2$\zsun. The assumed model star-formation histories assume a \citet{Chabrier03} initial mass function (IMF) and are exponentially declining with superimposed random starbursts with priors as described in \citet{Kauffmann03a}. We use the median of the PDF to estimate the
stellar mass and the 16th and 84th percentiles to estimate the associated uncertainty.
We calculate our own stellar masses instead of using those of \citet{Taylor11} for consistency with the VIPERS stellar masses. When comparing our stellar mass estimates with those of \citet{Taylor11} we see an offset which changes with redshift; there is a $0.1$\,dex offset at $z=0.05$ and $-0.05$\,dex offset at $z=0.35$. This is likely due to differences in the dust models and star-formation histories used in the spectral energy distribution (SED) fitting (see \citet{Wright17} who saw similar offsets between the \citet{Taylor11} and their stellar masses as a function of redshift). We find good agreement between our stellar masses and those derived using the {\sc MAGPHYS} code in \citet{Wright17}, with only a small 0.05\,dex offset at $z=0.35$. This offset is likely because \citet{Wright17} use observed frame $FUV-500\mu m$ data to estimate the stellar
masses and we only use $FUV-K$ magnitudes. As galaxies are dustier at high redshift this can cause a slight shift in the stellar masses.
We also compare our stellar masses to those in the MPA-JHU catalogue{\footnote{http://wwwmpa.mpa-garching.mpg.de/SDSS/DR7/}} which are calculated using fits to the SDSS DR7 $ugriz$ photometry. There is a systematic offset of 0.1\,dex as a result of using different stellar population models but we do not see any trend with redshift. Performing our analysis using the \citet{Taylor11} stellar mass measurements does not change our conclusions. In Appendix~\ref{sec:MF_compare} we compare our mass functions to those in the literature and generally find excellent agreement for both the GAMA and VIPERS samples.

\subsection{Incompleteness corrections}
We correct our number densities and mass functions for volume effects using the standard \Vmax method \citep{Schmidt68}, which weights the volume, V (the volume out to the redshift of each galaxy) by 1/\Vmax, which is the maximum volume over which a galaxy is visible in a magnitude limited survey or the upper redshift limit of a given redshift bin.
It is important to account for the variety of SED shapes in a sample \citep{Ilbert04}, as galaxies with a particular SED shape are visible out to different distances.
We do this by using the best-fitting SED model found when calculating the stellar masses. We scale the best-fitting model to the observed galaxy brightness and then shift it in redshift to determine the maximum distance out to which the galaxy could be seen, given the survey magnitude limits.

As we only select spectra which have a high enough SNR to reliably compute spectral indices, we must correct for the
fraction of missing galaxies before calculating number densities.
Following \citet{Wild09} we define the Quality Sampling Rate (QSR) as the fraction of galaxies above the SNR
threshold of 6.5, relative to the total number of galaxies in each stellar mass bin.
We compute the weight $w_i^\mathrm{QSR}$ in stellar mass bins of width 0.1\,dex and redshift bins with widths of $\delta z = 0.05-0.09$ for the GAMA survey, and $\delta z = 0.15-0.2$ for the VIPERS data. We also multiply the QSR correction by a factor to account for the fraction of spectra which do not have a PCA measurement due to e.g failure of the projection due to having $>20\%$ bad pixels, which is $<1\%$ of the total sample. The number of spectra that are missing due to fibre fringing, low quality redshifts ($nQ<3$) or highly uncertain PCA results is 13--33\%, depending on the redshift bin, and we account for this loss in our weighting scheme. We additionally account for the 5\% of spectra excluded from our sample which are not SDSS or GAMA spectra.

In VIPERS only $\sim40\%$ of the targets meeting the selection criteria in a given field were observed. We apply a statistical weight $w_i^\mathrm{TSR}$ as detailed in \citet{Guzzo14} to correct for the fraction of photometric objects which were not targeted (the target sampling rate, TSR). In GAMA the spectroscopic completeness is 98\%. To correct for the missing spectra we use a TSR correction of 0.98.
The ability to securely measure a spectroscopic redshift is a function of the observing conditions, and the brightness of the target. We correct for the fraction of targeted galaxies without secure redshifts (the spectroscopic sampling rate, SSR), and perform a completeness correction due to the colour selection (the colour sampling rate, CSR). Details of the SSR and CSR are given in \citet{Guzzo14} and \citet{Garilli14}. The GAMA sample has no colour selection criteria, so there are no SSR or CSR corrections to the low redshift sample.

The weight given to each galaxy ($w_i$) is

\begin{equation}
\frac{1} {\mathrm{V_{max}} \times w_i^\mathrm{SSR} \times w_i^\mathrm{TSR} \times w_i^\mathrm{CSR}  \times w_i^\mathrm{QSR}}.
\end{equation}

In Appendix~\ref{sec:MF_compare} we show that our corrections account for all sources of incompleteness as they allow us to recover total stellar mass functions which are consistent with published studies.

\subsection{Mass completeness limits}
\label{sec:mass_completeness}
The 90\% mass completeness limits were calculated in each redshift bin and separately for each spectral class following \citet{Pozzetti10}. It is important to do this separately for each spectral class, as our star-forming galaxy sample is complete to lower stellar masses than quiescent galaxies in a given redshift bin. We calculate the mass completeness limit using the stellar mass of each galaxy if it had a magnitude equal to the survey magnitude limit, so that log$_{10}(\mathrm{M_{lim}}) = \mathrm{log_{10}(M)} + 0.4(m-m_\mathrm{lim})$, where M is the galaxy stellar mass, $m$ is the observed apparent magnitude in the survey selection band ($r$ for GAMA, $i$ for VIPERS), and $m_\mathrm{lim}$ is the survey magnitude limit ($r=19.8$ mag for GAMA, $i=22.5$ mag for VIPERS). We use the $\mathrm{M_{lim}}$ of the faintest 20\% of these galaxies to represent galaxies with a typical M/L ratio near the survey limit. We then calculate the 90\% mass completeness limit of these typical faint galaxies, assuming $\mathrm{M_{lim}}$ is for galaxies in a relatively narrow redshift bin. The mass completeness limits for each spectral class and redshift bin are given in Table~\ref{tab:MF_fits}.

\begin{table*}
\centering
\caption{\label{tab:MF_fits} The single Schechter function fit parameters fitted to the total, PSB, red, green valley and star-forming mass functions in each redshift bin. The third column shows the total number of galaxies in each class in the redshift bin. The fourth column shows the 90\% mass completeness limit in log$_{10}(\mstar/\msun)$ for each bin and spectroscopic class. The fifth column shows the number of galaxies in each class in the redshift bin above the mass completeness limit. Uncertainties on each parameter account for the formal fitting errors on the Schechter function, uncertainty on the stellar masses, and cosmic variance. For the green valley galaxies in the highest redshift bin, and the PSBs in the lowest and highest redshift bins we fix $\alpha$ to -1.0 because there are not enough points to adequately constrain the faint end slope.}
  \begin{tabular}{cccccccc}\hline
Class & Redshift & Number & Compl. lim.& Number & Log$_{10}(\phi^\ast$/Mpc$^{-3})$ &  Log$_{10}(\mstar/\msun)$ & $\alpha$ \\
\hline
  Total & $0.05<z< 0.14$ &        13696 & 10.02 &        12303 & $3.87\pm0.40\times10^{-3}$ & $10.80 \pm  0.04$ & $-0.76 \pm  0.10$ \\
Total & $0.14<z< 0.21$ &        22321 & 10.36 &        14756 & $2.61\pm0.32\times10^{-3}$ & $10.93 \pm  0.04$ & $-1.02 \pm  0.11$ \\
Total & $0.21<z< 0.26$ &        13805 & 10.59 &         8110 & $1.82\pm0.30\times10^{-3}$ & $10.97 \pm  0.05$ & $-1.12 \pm  0.17$ \\
Total & $0.26<z< 0.35$ &        20846 & 10.87 &         9946 & $1.71\pm0.22\times10^{-3}$ & $11.02 \pm  0.05$ & $-1.07 \pm  0.23$ \\

  Total & $0.50<z< 0.65$ &        14091 & 10.33 &         5988 & $3.23\pm0.22\times10^{-3}$ & $10.77 \pm  0.04$ & $-0.48 \pm  0.14$ \\
Total & $0.65<z< 0.80$ &        13962 & 10.60 &         5037 & $3.12\pm0.17\times10^{-3}$ & $10.78 \pm  0.05$ & $-0.38 \pm  0.26$ \\
Total & $0.80<z< 1.00$ &        10125 & 10.87 &         2688 & $2.18\pm0.24\times10^{-3}$ & $10.86 \pm  0.09$ & $-0.73 \pm  0.50$ \\

  \hline
  PSB & $0.05<z<0.26$ &          172 & 10.57 &           33 & $7.62\pm16.61\times10^{-5}$ & $10.23 \pm  0.27$ & $-1.00 \pm  0.00$ \\

  PSB & $0.50<z<0.65$ &          180 & 10.45 &           73 & $5.04\pm6.88\times10^{-5}$ & $10.81 \pm  0.53$ & $-0.99 \pm  1.50$ \\
PSB & $0.65<z<0.80$ &          332 & 10.46 &          171 & $1.52\pm0.31\times10^{-4}$ & $10.58 \pm  0.23$ & $-0.37 \pm  1.07$ \\
PSB & $0.80<z<1.00$ &          362 & 10.81 &          121 & $1.32\pm0.68\times10^{-4}$ & $10.86 \pm  0.11$ & $-1.00 \pm  0.00$ \\

  \hline
  Quiescent & $0.05<z<0.14$ &         6936 & 10.00 &         6750 & $3.02\pm0.12\times10^{-3}$ & $10.65 \pm  0.03$ & $ 0.16 \pm  0.09$ \\
Quiescent & $0.14<z<0.21$ &         9655 & 10.39 &         8383 & $2.14\pm0.10\times10^{-3}$ & $10.80 \pm  0.03$ & $-0.20 \pm  0.12$ \\
Quiescent & $0.21<z<0.26$ &         5457 & 10.64 &         4365 & $1.37\pm0.09\times10^{-3}$ & $10.87 \pm  0.05$ & $-0.39 \pm  0.24$ \\
Quiescent & $0.26<z<0.35$ &         8921 & 10.94 &         5823 & $1.18\pm0.13\times10^{-3}$ & $10.90 \pm  0.06$ & $-0.15 \pm  0.36$ \\

  Quiescent & $0.50<z<0.65$ &         2829 & 10.36 &         2648 & $1.64\pm0.08\times10^{-3}$ & $10.68 \pm  0.04$ & $ 0.34 \pm  0.17$ \\
Quiescent & $0.65<z<0.80$ &         2493 & 10.62 &         2160 & $1.33\pm0.14\times10^{-3}$ & $10.73 \pm  0.05$ & $ 0.32 \pm  0.30$ \\
Quiescent & $0.80<z<1.00$ &         1192 & 11.01 &          712 & $0.81\pm0.77\times10^{-3}$ & $11.10 \pm  0.25$ & $-1.79 \pm  1.08$ \\

  \hline
  Green & $0.05<z<0.14$ &         2066 &  9.99 &         1926 & $1.04\pm0.06\times10^{-3}$ & $10.33 \pm  0.06$ & $ 0.21 \pm  0.26$ \\
Green & $0.14<z<0.21$ &         3351 & 10.38 &         2302 & $0.90\pm0.19\times10^{-4}$ & $10.42 \pm  0.08$ & $ 0.16 \pm  0.53$ \\
Green & $0.21<z<0.26$ &         2005 & 10.60 &         1209 & $0.66\pm0.12\times10^{-4}$ & $10.57 \pm  0.11$ & $-0.24 \pm  0.70$ \\
Green & $0.26<z<0.35$ &         3039 & 10.89 &         1334 & $0.68\pm0.24\times10^{-4}$ & $10.70 \pm  0.19$ & $-0.65 \pm  1.39$ \\

  Green & $0.50<z<0.65$ &          960 & 10.36 &          792 & $6.64\pm0.57\times10^{-4}$ & $10.63 \pm  0.10$ & $-0.24 \pm  0.46$ \\
Green & $0.65<z<0.80$ &          868 & 10.68 &          589 & $4.54\pm2.99\times10^{-4}$ & $10.58 \pm  0.15$ & $ 0.51 \pm  1.11$ \\
Green & $0.80<z<1.00$ &          476 & 10.95 &          251 & $6.35\pm3.66\times10^{-4}$ & $10.80 \pm  0.09$ & $-1.00 \pm  0.00$ \\

  \hline
   SF & $0.05<z<0.14$ &         4661 & 10.10 &         3060 & $2.33\pm0.35\times10^{-3}$ & $10.35 \pm  0.07$ & $-0.75 \pm  0.28$ \\
SF & $0.14<z<0.21$ &         9145 & 10.34 &         3999 & $1.57\pm0.41\times10^{-3}$ & $10.57 \pm  0.09$ & $-1.26 \pm  0.34$ \\
SF & $0.21<z<0.26$ &         6243 & 10.54 &         2440 & $1.23\pm0.48\times10^{-3}$ & $10.68 \pm  0.12$ & $-1.48 \pm  0.52$ \\
SF & $0.26<z<0.35$ &         8712 & 10.80 &         2466 & $1.28\pm0.52\times10^{-3}$ & $10.73 \pm  0.15$ & $-1.51 \pm  0.75$ \\

   SF & $0.50<z<0.65$ &         3437 & 10.17 &         2150 & $1.65\pm0.24\times10^{-3}$ & $10.52 \pm  0.06$ & $-0.83 \pm  0.21$ \\
SF & $0.65<z<0.80$ &         4388 & 10.43 &         1919 & $1.56\pm0.25\times10^{-3}$ & $10.64 \pm  0.08$ & $-0.83 \pm  0.34$ \\
SF & $0.80<z<1.00$ &         2984 & 10.68 &          909 & $1.01\pm0.11\times10^{-3}$ & $10.66 \pm  0.11$ & $-0.42 \pm  0.63$ \\

  \hline
  \end{tabular}
\end{table*}

\subsection{Uncertainties}\label{sec:uncertainties}
The total uncertainty in the number and stellar mass densities are calculated by adding in quadrature the errors due to sample size, uncertainty on the stellar masses, and those due to cosmic variance.
The errors due to sample size (i.e. Poisson uncertainty) are estimated following \citet{Moustakas13}, where the method of \citet{Gehrels86} is used to compute the upper and lower limits on the uncertainty in the mass function. This method properly accounts for the uncertainty on a value when there are a small number of galaxies per mass bin, which is common at the high mass end of the stellar mass function.
We estimate the cosmic variance in each GAMA and VIPERS field with the publicly available tool {\sc getcv} \citep{Moster11}. As the GAMA survey covers three separate fields, and VIPERS covers two separate fields, the uncertainty due to cosmic variance is reduced further, as the uncertainties for each field are combined following \citet{Moster11}. The uncertainties due to cosmic variance are minimised by the large survey volumes, and
range from 5--11\% at \mstar$\sim10^{10.6}$ and 6--12\% at \mstar$>10^{11}$ in the GAMA survey, and 4--6\% at \mstar$\sim10^{10.6}$ to 6--8\% at \mstar$>10^{11}$ in the VIPERS survey.
To estimate the impact of the uncertainty in stellar mass on the mass function and the cumulative number densities, we perturb each stellar mass by a random amount drawn from a Gaussian distribution with a standard deviation equal to the $1\sigma$ error on the stellar mass. We do this for 100 realisations and take the standard deviation of the number and mass density in each stellar mass and redshift bin.

There are also systematic uncertainties in the stellar mass due to the choice of stellar population models and IMF of around $\sim0.2-0.3$\,dex; see \citet{Wright17} for a discussion of the effect of different stellar mass estimates on the galaxy stellar mass function. We note that we have used exactly the same method to calculate the stellar mass in both samples, which is crucial to make this comparison valid, therefore systematics between the two samples are minimised.

\subsection{Aperture bias}
\label{sec:Aperture_bias}
We note that our results could be affected by aperture bias, as the fibre spectra cover a larger proportion of the galaxy at high redshift. Galaxy outskirts are usually bluer than the centre in spiral galaxies, but early-type galaxies tend to have flat or positive colour gradients \citep{Gonzalez-Perez11}. Since the majority of the quiescent population is likely comprised of early-type galaxies, aperture bias should have a negligible impact on our results involving the quiescent population.  Indeed, both the SDSS and GAMA fibres cover $>90\%$ of the flux from a model galaxy with an effective radius of 4\,kpc and with S\'{e}rsic index of 4 at $z>0.1$. However, for the star-forming, post-starburst and green-valley populations, at low redshift these may be classified as having redder spectra. We may therefore select fewer galaxies at low redshift than at high redshift, leading to an overestimation of the decline in these populations with time. The SDSS (GAMA) fibres cover $32-78\%$ ($18-56\%$) of the flux from a model galaxy with S\'{e}rsic index of 1 at $z=0.1-0.3$, respectively. We note that green valley and PSB galaxies have a range of S\'{e}rsic indices and so will not be as affected by aperture bias as the star-forming galaxies, which tend to have lower S\'{e}rsic indices.
Furthermore, \citet{Pracy12} found that low redshift PSBs showed declining Balmer absorption line strengths with increasing radius. At higher redshift we may select fewer PSBs because aperture effects dilute the Balmer line strength, but the amount by which aperture bias affects the spectra of low redshift PSBs may not be equal to the amount by which Balmer absorption bias affects the spectra of high redshift PSBs.
We test the effects of aperture bias on the classifications of galaxies in Figure~\ref{fig:UVJ} and find no evidence that the broad band colours of spectroscopically classified galaxies change with redshift. We therefore conclude that aperture bias has a negligible effect on our results.

Whilst aperture corrections are available for physical parameters such as SFR and have been shown to be robust for large galaxy populations \citep{Brough13, Richards16}, aperture corrections for detailed stellar population analysis are not available. This issue will be addressed by next generation integral field spectroscopic surveys such as Mapping Nearby Galaxies at APO (MaNGA, \citealt{Bundy15}) and Sydney-Australian-Astronomical-Observatory Multi-object Integral-Field Spectrograph (SAMI, \citealt{Croom12, Bryant15}).

\section{Results}
\label{sec:Results}
Most previous studies of the build-up of the number and stellar mass density of star-forming and quiescent galaxies have used broad-band data \citep[e.g.][]{Arnouts07, Ilbert13, Muzzin13, Moustakas13}. Studies of the actively quenching populations which may be responsible for the build-up of the quiescent population have been limited to small samples of spectroscopically identified PSBs ($<20$) at $0.5<z<1$ \citep{Wild09, Vergani10}, which cannot be split by mass due to small number statistics (although see \citealt{Pattarakijwanich16} who select $\sim6000$ PSBs from the SDSS at $0.05<z<1.3$). We use our spectroscopic classifications of a large sample of quiescent, star-forming, green-valley and PSB galaxies to see how each of the populations change as a function of stellar mass over a wide redshift range from $0.05<z<1.0$. In the following analysis we only use redshift bins above the 90\% mass completeness limit.

\subsection{Fractions in each spectral class}
\begin{figure}
\begin{center}
\includegraphics[width=0.48\textwidth]{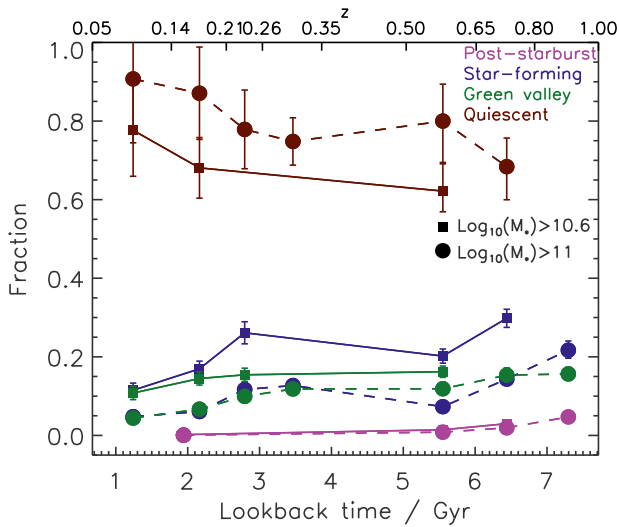}
\end{center}
\caption{The volume-corrected fraction of quiescent population, star-forming, green-valley and PSB galaxies as a function of stellar mass and redshift. Intermediate mass galaxies with $\mstar>10^{10.6}$\msun are shown as small squares joined with solid lines, and high mass galaxies with $\mstar>10^{11}$\msun are shown as large circles joined with dashed lines. Uncertainties are propagated from the number densities and include Poisson errors, cosmic variance and those due to uncertainties on the stellar mass, which are typically smaller than the symbol size.}
\label{fig:passive_frac}
\end{figure}

In Figure~\ref{fig:passive_frac} we show the fraction of quiescent, star-forming, green-valley and PSB galaxies in each mass bin as a function of redshift. As a function of stellar mass, $>60\%$ of intermediate mass ($\mstar>10^{10.6}$\msun; small squares), and $>70\%$ of high mass ($\mstar>10^{11}$\msun; large circles) galaxies are in the quiescent population, and high mass star-forming galaxies are rare ($<30\%$ and $<25\%$ for intermediate and high mass galaxies, respectively).
The quiescent fraction increases from high to low redshift, while the star-forming fraction is decreasing towards low redshift, independent of stellar mass. The fraction of galaxies in the green valley at intermediate masses ($\mstar>10^{10.6}$\msun; small squares) is $\sim16\%$ at $z=0.6$, decreasing slightly to 11\% at $z=0.1$. The fraction of high mass green valley galaxies decreases more steeply from 15\% to 4\% from $z=0.7$ to $z=0.1$. The PSB galaxies are rare at any redshift, and comprise only $0.06-0.2$\% of the total galaxy population at $z<0.35$, rising to $2-3$\% of the population at $z\sim0.7$, depending on stellar mass.
Our rising PSB fraction with redshift is in agreement with the findings of \citet{Dressler13} and \citet{Wild09, Wild16}. Our results also agree with the lower limits on the number density of compact, massive ($M_\ast>10^{11} M_\odot$) E+A galaxies at $0.2<z<0.8$ from \citet{Zahid16}.
The PSBs are rare compared to green valley galaxies, this may be because they are only visible for a short time, whereas galaxies in the green valley may spend longer there, as we discuss in the following sections.

\subsection{Number density evolution}
\label{sec:NumberDensities}
\begin{figure}
\begin{center}
\includegraphics[scale=0.7]{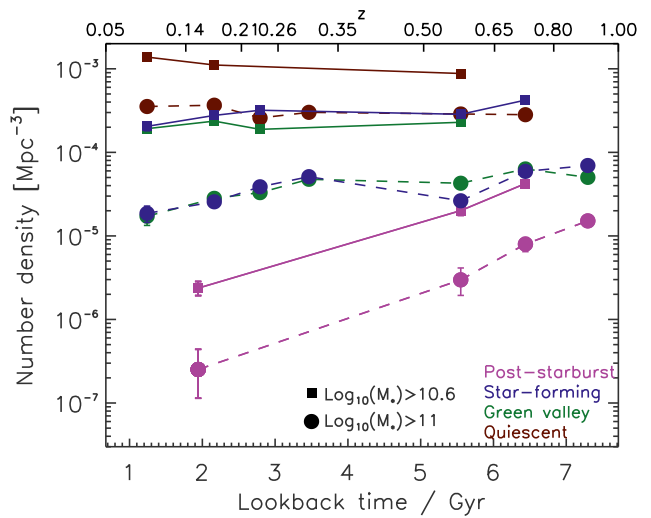}
\end{center}
\caption{The completeness corrected comoving number densities (Mpc$^{-3}$) for star-forming, red, green valley and PSB galaxies in redshift and stellar bins. Intermediate mass galaxies with $\mstar>10^{10.6}$\msun are shown as small squares joined with solid lines, and high mass galaxies with $\mstar>10^{11}$\msun are shown as large circles joined with dashed lines. Points are only shown for bins above the 90\% completeness limit. Errors include uncertainties due to sample size and cosmic variance.}
\label{fig:numberdens}
\end{figure}

We present the completeness corrected cumulative number densities of spectroscopically identified star-forming, quiescent, green valley and PSB galaxies as a function of redshift and mass in Table~\ref{tab:NumberDensities} and Figure~\ref{fig:numberdens}. If we move the spectral classification boundaries by $\Delta{\rm (PC)}=0.1$ (twice as large as the typical uncertainty on PC1 and PC2) then the quiescent, star-forming and green valley number densities change a negligible amount, but the PSB number densities change by a factor of two. The trends that we observe with redshift remain unchanged. Qualitatively our results are robust to changes in the spectral classification boundaries and stellar mass binning.

\begin{table*}
\centering
\caption{The cumulative comoving log number densities (Mpc$^{-3}$) of galaxies in each spectroscopic class in redshift bins above a given stellar mass limit. Values are only shown for bins above the 90\% mass completeness limit in each spectroscopic class. Uncertainties include Poisson errors, cosmic variance and those due to uncertainties on the stellar mass.}
\begin{tabular}{cccccc}\hline
Redshift & All & Star forming & PSB & Quiescent & Green \\
\hline
 & & log$_{10}($M/\mstar$)>10.6$ & & \\
\hline
 $0.05 < z < 0.14$ & $-2.75^{+ 0.04}_{- 0.05}$ & $-3.69^{+ 0.05}_{- 0.05}$ & $-5.63^{+ 0.08}_{- 0.08}$ & $-2.86^{+ 0.04}_{- 0.05}$ & $-3.72^{+ 0.05}_{- 0.05}$ \\
 $0.14 < z < 0.21$ & $-2.79^{+ 0.03}_{- 0.04}$ & $-3.56^{+ 0.03}_{- 0.04}$ & $-5.63^{+ 0.08}_{- 0.08}$ & $-2.95^{+ 0.03}_{- 0.04}$ & $-3.63^{+ 0.03}_{- 0.04}$ \\
 $0.21 < z < 0.26$ & $-2.91^{+ 0.03}_{- 0.03}$ & $-3.50^{+ 0.03}_{- 0.03}$ & $-5.63^{+ 0.08}_{- 0.08}$ & - & $-3.72^{+ 0.03}_{- 0.04}$ \\
 $0.26 < z < 0.35$ & - & - & - & - & - \\
 $0.50 < z < 0.65$ & $-2.85^{+ 0.03}_{- 0.03}$ & $-3.54^{+ 0.03}_{- 0.03}$ & $-4.70^{+ 0.05}_{- 0.06}$ & $-3.06^{+ 0.03}_{- 0.03}$ & $-3.64^{+ 0.03}_{- 0.03}$ \\
 $0.65 < z < 0.80$ & $-2.85^{+ 0.02}_{- 0.02}$ & $-3.37^{+ 0.02}_{- 0.03}$ & $-4.37^{+ 0.04}_{- 0.04}$ & - & - \\
 $0.80 < z < 1.00$ & - & - & - & - & - \\
\hline
 & & log$_{10}($M/\mstar$)>11$ & & \\
\hline
 $0.05 < z < 0.14$ & $-3.41^{+ 0.05}_{- 0.06}$ & $-4.73^{+ 0.09}_{- 0.10}$ & $-6.60^{+ 0.24}_{- 0.34}$ & $-3.45^{+ 0.05}_{- 0.06}$ & $-4.76^{+ 0.09}_{- 0.10}$ \\
 $0.14 < z < 0.21$ & $-3.38^{+ 0.04}_{- 0.04}$ & $-4.59^{+ 0.05}_{- 0.06}$ & $-6.60^{+ 0.24}_{- 0.34}$ & $-3.44^{+ 0.04}_{- 0.04}$ & $-4.55^{+ 0.05}_{- 0.06}$ \\
 $0.21 < z < 0.26$ & $-3.48^{+ 0.04}_{- 0.04}$ & $-4.41^{+ 0.05}_{- 0.05}$ & $-6.60^{+ 0.24}_{- 0.34}$ & $-3.59^{+ 0.04}_{- 0.04}$ & $-4.48^{+ 0.05}_{- 0.05}$ \\
 $0.26 < z < 0.35$ & $-3.39^{+ 0.02}_{- 0.03}$ & $-4.29^{+ 0.03}_{- 0.03}$ & $-6.60^{+ 0.24}_{- 0.34}$ & $-3.52^{+ 0.02}_{- 0.03}$ & $-4.32^{+ 0.03}_{- 0.03}$ \\
 $0.50 < z < 0.65$ & $-3.44^{+ 0.03}_{- 0.04}$ & $-4.58^{+ 0.06}_{- 0.07}$ & $-5.53^{+ 0.14}_{- 0.19}$ & $-3.54^{+ 0.03}_{- 0.04}$ & $-4.37^{+ 0.05}_{- 0.05}$ \\
 $0.65 < z < 0.80$ & $-3.38^{+ 0.03}_{- 0.03}$ & $-4.23^{+ 0.04}_{- 0.05}$ & $-5.10^{+ 0.08}_{- 0.10}$ & $-3.55^{+ 0.03}_{- 0.03}$ & $-4.20^{+ 0.04}_{- 0.04}$ \\
 $0.80 < z < 1.00$ & $-3.49^{+ 0.03}_{- 0.03}$ & $-4.16^{+ 0.04}_{- 0.04}$ & $-4.82^{+ 0.05}_{- 0.06}$ & - & $-4.30^{+ 0.04}_{- 0.04}$ \\

\hline
\end{tabular}
\label{tab:NumberDensities}
\end{table*}

\begin{table*}
\centering
\caption{The cumulative comoving log mass densities (\msun Mpc$^{-3}$) of galaxies in each spectroscopic class in redshift bins above a given stellar mass limit. Values are only shown for bins above the 90\% mass completeness limit in each spectroscopic class. Uncertainties include Poisson errors, cosmic variance and those due to uncertainties on the stellar mass.}
\begin{tabular}{cccccc}\hline
Redshift & All & Star forming & PSB & Quiescent & Green \\
\hline
 & & log$_{10}($M/\mstar$)>10.6$ & & \\
\hline
 $0.05 < z < 0.14$ & $ 8.17^{+ 0.04}_{- 0.05}$ & $ 7.15^{+ 0.05}_{- 0.06}$ & $ 5.30^{+ 0.09}_{- 0.11}$ & $ 8.08^{+ 0.04}_{- 0.05}$ & $ 7.09^{+ 0.05}_{- 0.05}$ \\
 $0.14 < z < 0.21$ & $ 8.16^{+ 0.03}_{- 0.04}$ & $ 7.28^{+ 0.04}_{- 0.04}$ & $ 5.30^{+ 0.09}_{- 0.11}$ & $ 8.03^{+ 0.03}_{- 0.04}$ & $ 7.22^{+ 0.03}_{- 0.04}$ \\
 $0.21 < z < 0.26$ & $ 8.04^{+ 0.03}_{- 0.03}$ & $ 7.36^{+ 0.03}_{- 0.03}$ & $ 5.30^{+ 0.09}_{- 0.11}$ & - & $ 7.16^{+ 0.03}_{- 0.04}$ \\
 $0.26 < z < 0.35$ & - & - & - & - & - \\
 $0.50 < z < 0.65$ & $ 8.09^{+ 0.03}_{- 0.03}$ & $ 7.27^{+ 0.03}_{- 0.03}$ & $ 6.15^{+ 0.05}_{- 0.06}$ & $ 7.93^{+ 0.03}_{- 0.03}$ & $ 7.25^{+ 0.03}_{- 0.03}$ \\
 $0.65 < z < 0.80$ & $ 8.11^{+ 0.02}_{- 0.02}$ & $ 7.47^{+ 0.02}_{- 0.03}$ & $ 6.50^{+ 0.04}_{- 0.04}$ & - & - \\
 $0.80 < z < 1.00$ & - & - & - & - & - \\
\hline
 & & log$_{10}($M/\mstar$)>11$ & & \\
\hline
 $0.05 < z < 0.14$ & $ 7.80^{+ 0.05}_{- 0.06}$ & $ 6.59^{+ 0.09}_{- 0.11}$ & $ 4.94^{+ 0.24}_{- 0.34}$ & $ 7.75^{+ 0.05}_{- 0.06}$ & $ 6.37^{+ 0.08}_{- 0.10}$ \\
 $0.14 < z < 0.21$ & $ 7.84^{+ 0.04}_{- 0.04}$ & $ 6.69^{+ 0.05}_{- 0.06}$ & $ 4.94^{+ 0.24}_{- 0.34}$ & $ 7.77^{+ 0.04}_{- 0.04}$ & $ 6.63^{+ 0.05}_{- 0.06}$ \\
 $0.21 < z < 0.26$ & $ 7.73^{+ 0.04}_{- 0.04}$ & $ 6.81^{+ 0.05}_{- 0.05}$ & $ 4.94^{+ 0.24}_{- 0.34}$ & $ 7.63^{+ 0.04}_{- 0.04}$ & $ 6.69^{+ 0.05}_{- 0.05}$ \\
 $0.26 < z < 0.35$ & $ 7.84^{+ 0.02}_{- 0.03}$ & $ 6.89^{+ 0.03}_{- 0.03}$ & $ 4.94^{+ 0.24}_{- 0.34}$ & $ 7.73^{+ 0.02}_{- 0.03}$ & $ 6.84^{+ 0.03}_{- 0.03}$ \\
 $0.50 < z < 0.65$ & $ 7.74^{+ 0.03}_{- 0.04}$ & $ 6.54^{+ 0.06}_{- 0.07}$ & $ 5.60^{+ 0.14}_{- 0.19}$ & $ 7.66^{+ 0.03}_{- 0.04}$ & $ 6.78^{+ 0.05}_{- 0.05}$ \\
 $0.65 < z < 0.80$ & $ 7.81^{+ 0.03}_{- 0.03}$ & $ 6.91^{+ 0.04}_{- 0.05}$ & $ 6.02^{+ 0.09}_{- 0.11}$ & $ 7.66^{+ 0.03}_{- 0.03}$ & $ 6.96^{+ 0.04}_{- 0.04}$ \\
 $0.80 < z < 1.00$ & $ 7.71^{+ 0.03}_{- 0.03}$ & $ 7.00^{+ 0.03}_{- 0.04}$ & $ 6.37^{+ 0.05}_{- 0.06}$ & - & $ 6.89^{+ 0.04}_{- 0.04}$ \\

\hline
\end{tabular}
\label{tab:MassDensities}
\end{table*}

For intermediate ($\mstar>10^{10.6}$\msun) and high mass ($\mstar>10^{11}$\msun) star-forming galaxies the population declines in number density between $z=0.6$ and $z=0.1$. For intermediate mass ($\mstar>10^{10.6}$\msun) quiescent galaxies the population grows in number density by a factor of 1.58 between $z=0.6$ and $z=0.1$. We find that the number density of high mass ($\mstar>10^{11}$\msun) quiescent galaxies increases by a factor of 1.23 between $z=0.6$ and $z=0.1$.
Our results are similar to those of \citet{Moustakas13}, who found that the number density of quiescent galaxies (selected using a cut in the broadband photometry-derived \mstar-SFR relation) grows more slowly for high mass galaxies from $z=1$ to $z=0.1$. We cannot recover the number density of less massive ($\mstar<10^{10}$\msun) galaxies beyond the lowest redshift bin as our sample becomes incomplete in low mass galaxies at $z>0.14$. Deeper spectroscopy or the use of photometric galaxy classification methods \citep{Wild14} are still required to probe the low stellar mass quiescent galaxy regime. It may be that the quiescent population is growing more rapidly at low redshift than suggested by our results, but only at lower masses than those probed by our study \citep{Tinker13, Muzzin13}.

To-date, there have been few studies of the number densities of candidate transition (green-valley and PSB) galaxies. We find that green valley galaxies with intermediate masses of $\mstar>10^{10.6}$\msun have an approximately flat number density of $\sim10^{-3.7}$Mpc$^{-3}$ from $z=1$ to $z=0$. At high stellar masses ($\mstar>10^{11}$\msun), the number density of green valley galaxies decreases by an order of magnitude from $z=1$ to $z=0$.

The number density of intermediate mass ($\mstar>10^{10.6}$\msun) and high mass ($\mstar>10^{11}$\msun) PSB galaxies decreases by an order of magnitude in the redshift range $0.2<z<0.6$. Our results are qualitatively consistent with the results of \citet{Wild16} who found a factor of three decrease in the number density of $\mstar>10^{10.6}$\msun PSBs from $z=2$ to $z=0.5$ (see also \citealt{Dressler13}). Using VIMOS-VLT Deep Survey (VVDS) spectra, \citet{Wild09} found that there are more galaxies passing through the PSB phase at high redshift than at low redshift. The number densities of intermediate mass PSBs in our study at $0.5<z<0.65$ are similar to those in \citet{Wild16}, who found a number density of $10^{-4.9}$ Mpc$^{-3}$ for $\mstar>10^{10.5}$\msun PSBs at $0.5<z<1$. However, the PSB number density at $0.65<z<0.8$ is larger that that of \citet{Wild16}. This discrepancy may be because \citet{Wild16} use a photometric selection method which may not be as sensitive to PSB features as the spectroscopic selection used in our study.
The number density of PSB galaxies identified spectroscopically
from the VVDS survey with $0.5<z<1.0$ by \citet{Wild09} was $10^{-4}$Mpc$^{-3}$ for galaxies with $\mstar>10^{9.75}$\msun, measured from
16 PSB galaxies. As we are highly incomplete at such low stellar masses we cannot directly compare to the results from \citet{Wild09}, which used VVDS data which is two magnitudes deeper than the VIPERS survey.

The stellar mass densities (Table~\ref{tab:MassDensities}) show very similar behaviour to the number densities. For intermediate mass ($\mstar>10^{10.6}$\msun) quiescent galaxies the population grows in stellar mass density by a factor of 1.41 between $z=0.6$ and $z=0.1$. The mass density of high mass ($\mstar>10^{11}$\msun) quiescent galaxies grows by a factor of 1.23 between $z=0.6$ and $z=0.1$. Our results for the growth of the quiescent population are smaller than those of \citet{Bell04}, \citet{Brown07}, and \citet{Arnouts07}, who found that the quiescent population has doubled in mass in the range $0<z<1$.
The differences between our measured mass growth rate and literature studies may be because our spectroscopic selection, stellar mass and redshift range are slightly different to those in other studies. Furthermore, we have checked that aperture bias does not cause us to misclassify large numbers of galaxies at low redshift, see Figure~\ref{fig:UVJ}. Our results show that, in general, there were more transition galaxies at high redshift than in the local Universe.

\subsection{Evolution of Mass functions}
\begin{figure*}
\begin{minipage}[t]{1.0\textwidth}
\begin{center}
$\begin{array}{cc}
\includegraphics[scale=0.7]{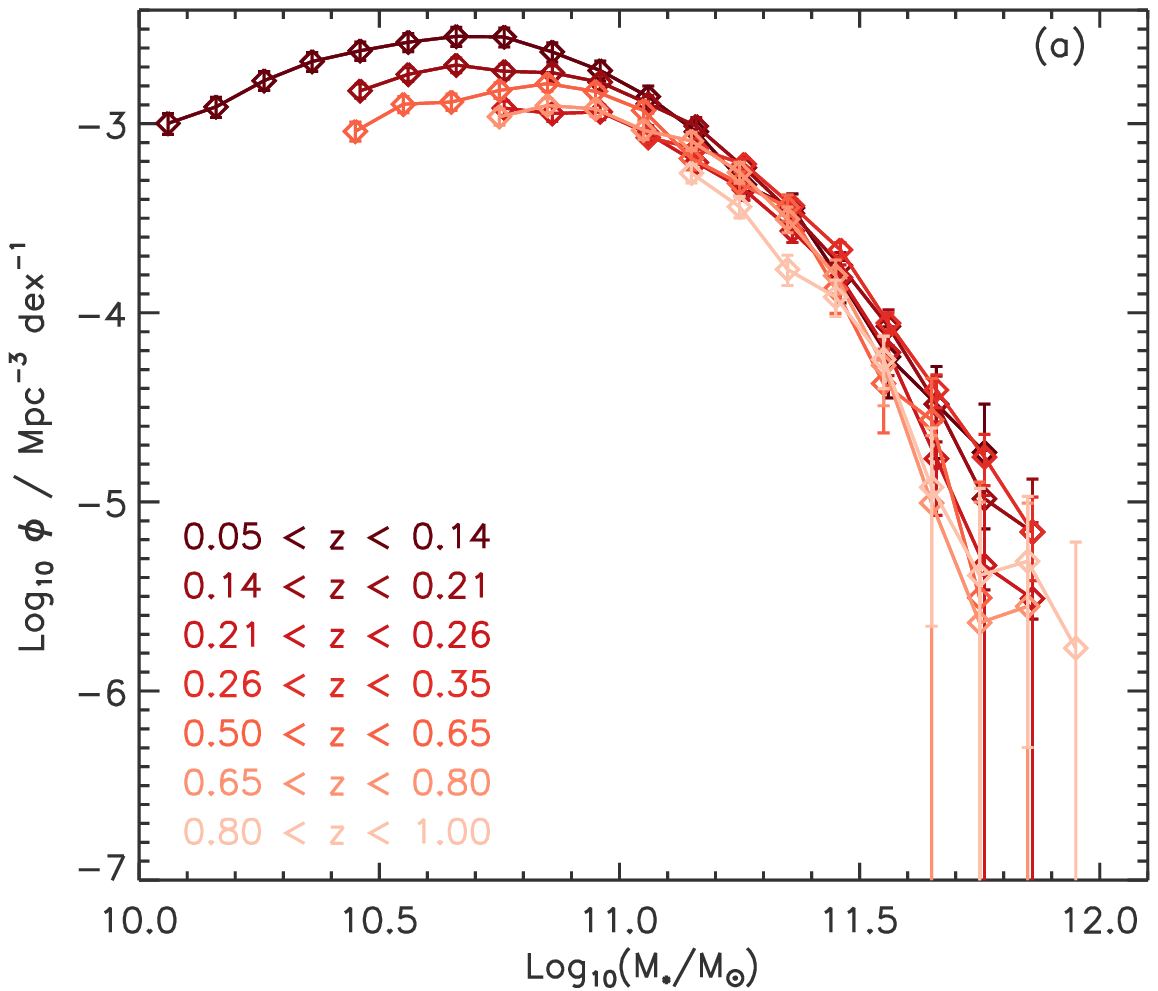} &
\includegraphics[scale=0.7]{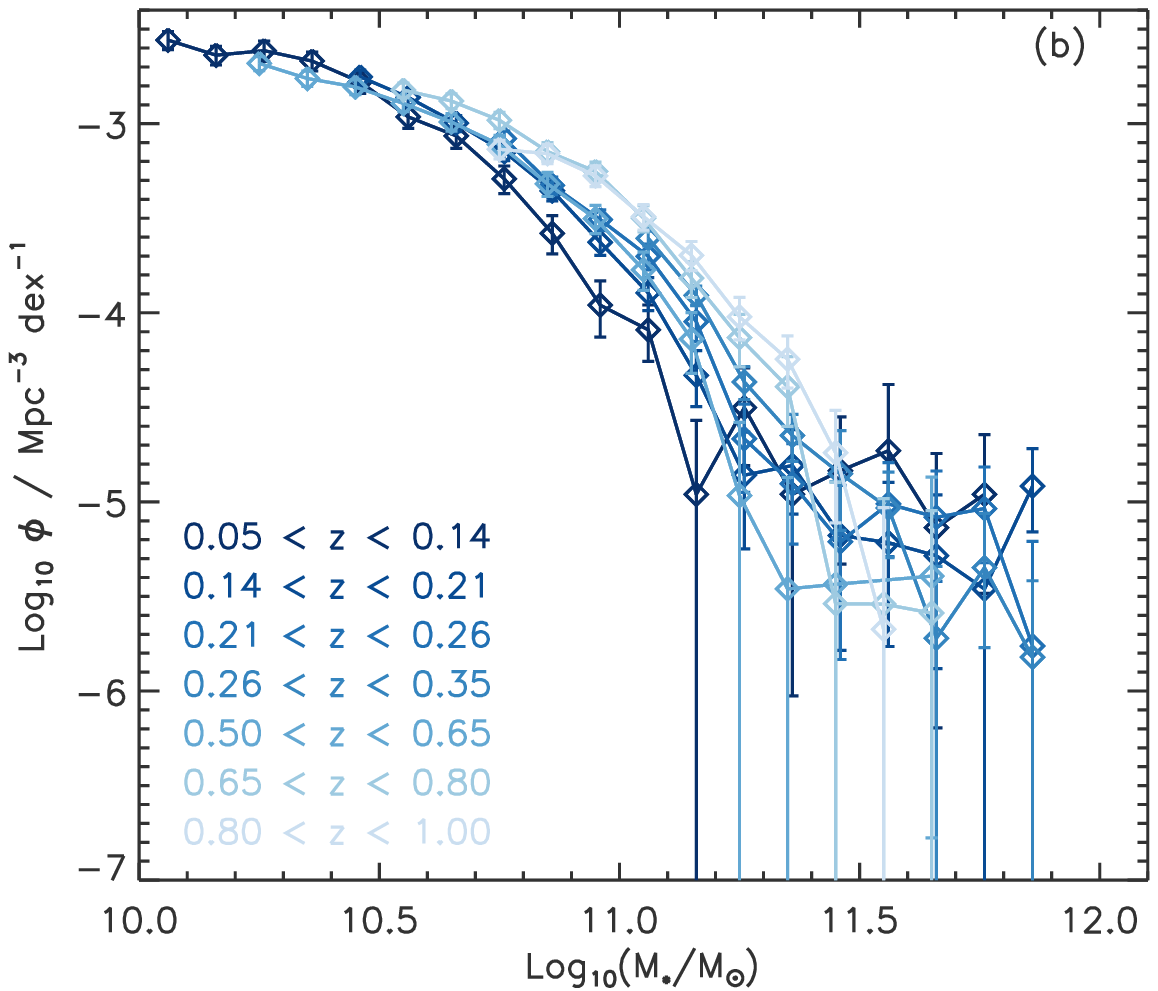} \\
\includegraphics[scale=0.7]{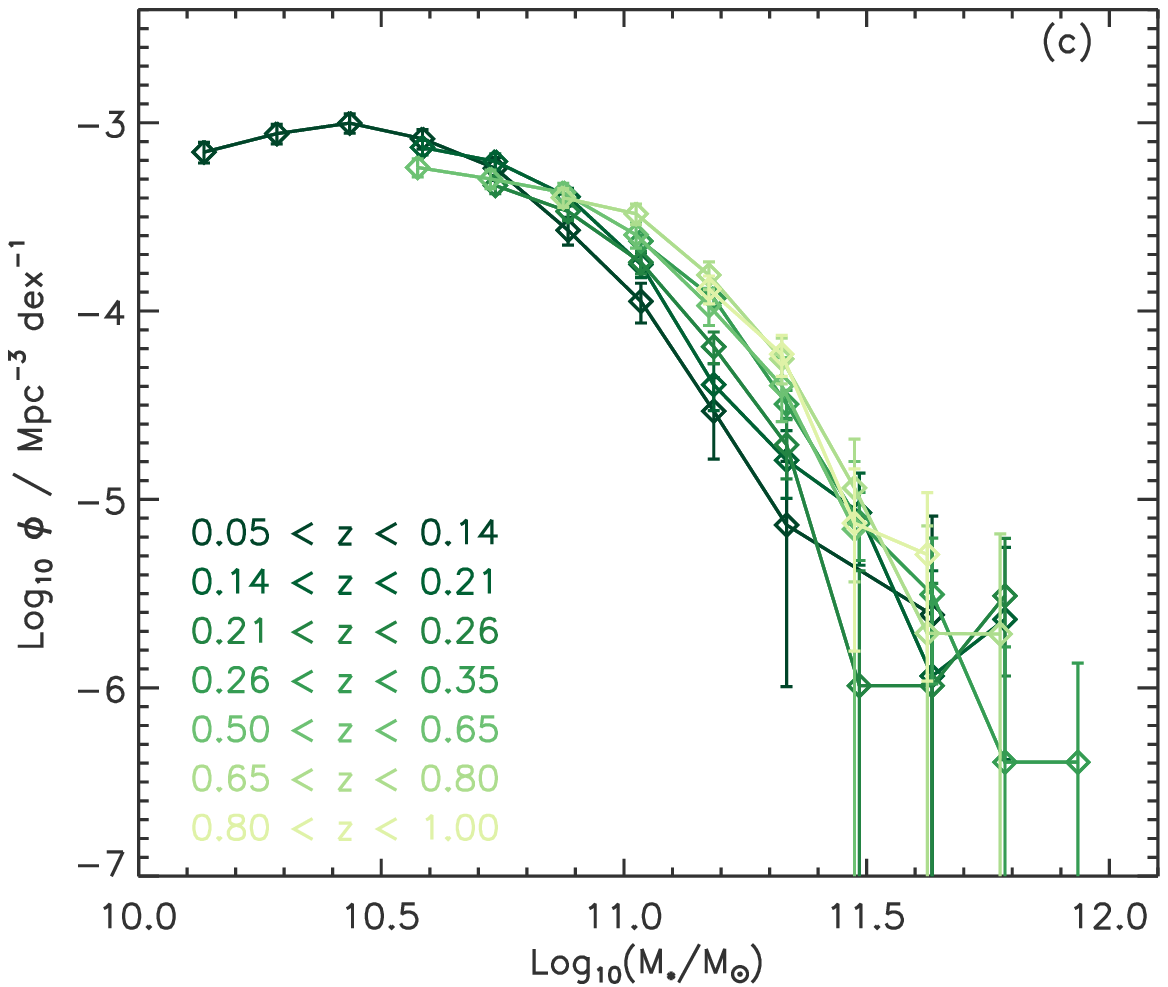} &
\includegraphics[scale=0.7]{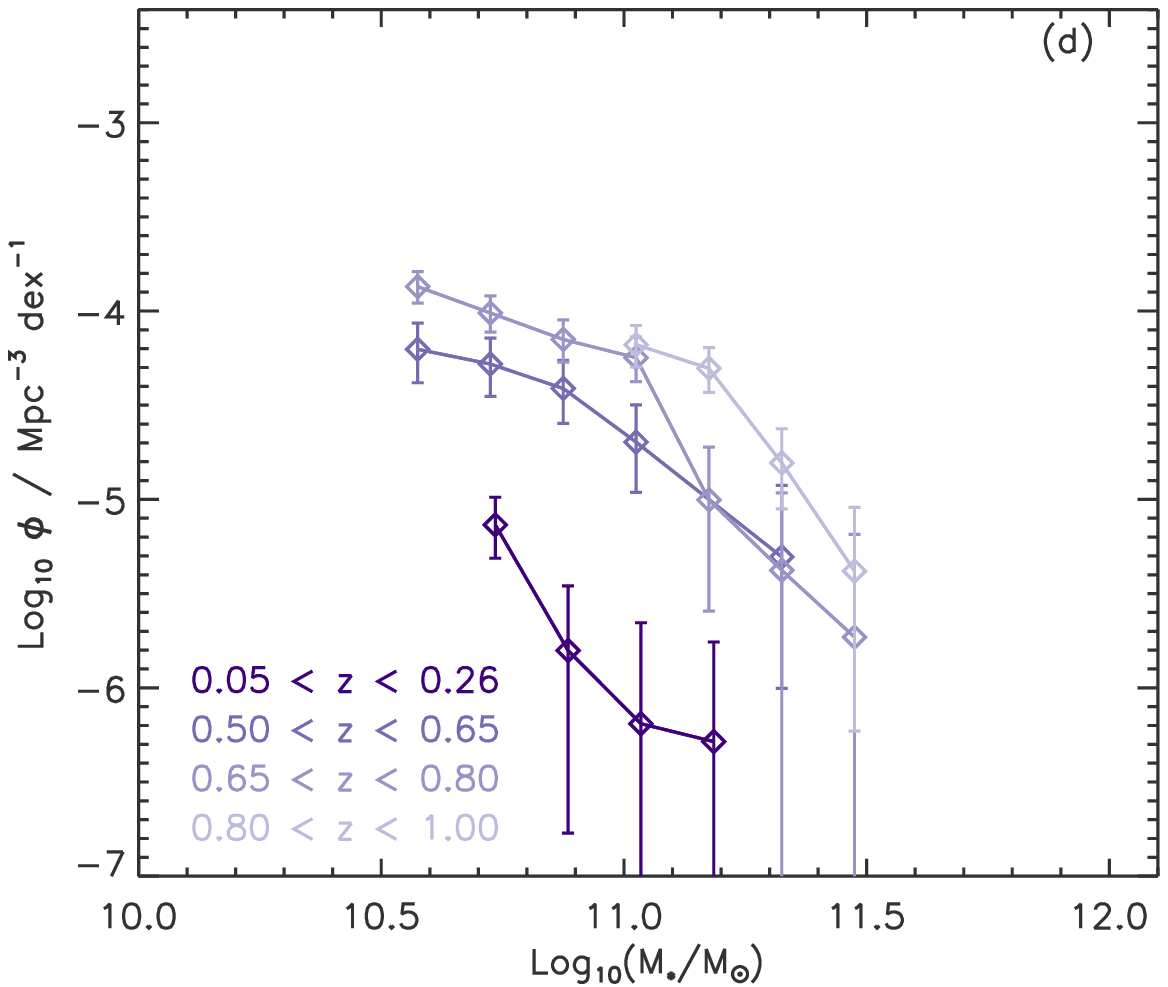} \\
\end{array}$
\end{center}
\end{minipage}
\caption{The stellar mass functions (corrected for incompleteness) for quiescent (a), star-forming (b), green-valley (c) and PSBs (d), in different redshift bins. Points are only shown where the bin is $>90\%$ complete in stellar mass. Errors include Poisson uncertainties, cosmic variance and from uncertainties on the stellar masses.}
\label{fig:massfunc_red_green}
\end{figure*}

We present the mass functions of the red, star-forming, green-valley and PSB galaxies in Figure~\ref{fig:massfunc_red_green}.
We fit our mass functions with single Schechter functions \citep{Schechter76}, using the {\sc {impro}} IDL library\footnote{https://github.com/moustakas/impro} \citep{Moustakas13}. We do not fit our mass functions with double Schechter functions as we do not see an upturn at low stellar masses. The Schechter fit parameters are in Table~\ref{tab:MF_fits}. The uncertainties on the number density ($\phi$) include contributions from Poisson errors, cosmic variance, and from uncertainties on the stellar mass estimated via a Monte Carlo method with 100 realisations (Section \ref{sec:uncertainties}).

The mass functions of the quiescent galaxies show a clear build-up in the low mass end from $z=1$ to $z=0$, and a smaller increase in the number density of high mass galaxies, as is commonly found in the literature. Conversely, the mass functions of the star-forming galaxies show that from $z=1$ to $z=0$ there is a decline in the number density of massive galaxies with redshift. Note that our spectroscopically defined quiescent population mass function is different to that selected on $u-r$ and optical colour from \citet{Baldry12} (also using GAMA data), as we find fewer low mass galaxies (see Appendix~\ref{sec:MF_compare}). This may be because \citet{Baldry12} separate star-forming and quiescent galaxies using broad-band colours, and we use a cleaner spectroscopic selection that likely has less contamination by dusty objects. Furthermore, we separate quiescent from green valley galaxies, whereas \citet{Baldry12} do not make this discrimination, meaning that green valley galaxies will be mixed with the red and blue populations defined with broad-band colours \citep{Taylor15}. See Appendix~\ref{sec:colourmag} for an analysis of the broad-band colours of galaxies in each spectral class.

As seen in Section~\ref{sec:NumberDensities}, the green valley galaxy mass functions show a negligible build-up at $\mstar \sim10^{10.5}\msun$ with redshift, but there is more evolution at the high mass end of the mass function. There were more high mass galaxies in the green valley at high redshift than at low redshift. The PSB mass function exhibits stronger evolution in the mass function than green valley galaxies, with galaxies in the PSB phase more massive at high redshift. Our transition galaxy mass functions are consistent with more massive galaxies quenching earlier, and less massive galaxies quenching later. Similar results were found by \citet{Goncalves12} for green valley galaxies.

\section{How fast do galaxies quench?}
Previous studies using broad-band photometry have not reached a consensus on the relative importance of the fast and slow channels for galaxy quenching and building of the quiescent galaxy population. Spectroscopic surveys offer the unique advantage of being able to cleanly identify both quiescent galaxies and candidate transition galaxies with different quenching time-scales. In this section we discuss our results in terms of the relative importance of these two channels for building the quiescent population.

\subsection{Quiescent galaxies}
\label{sec:quiescent}
\begin{figure}
\begin{center}
\includegraphics[width=0.48\textwidth, clip=true]{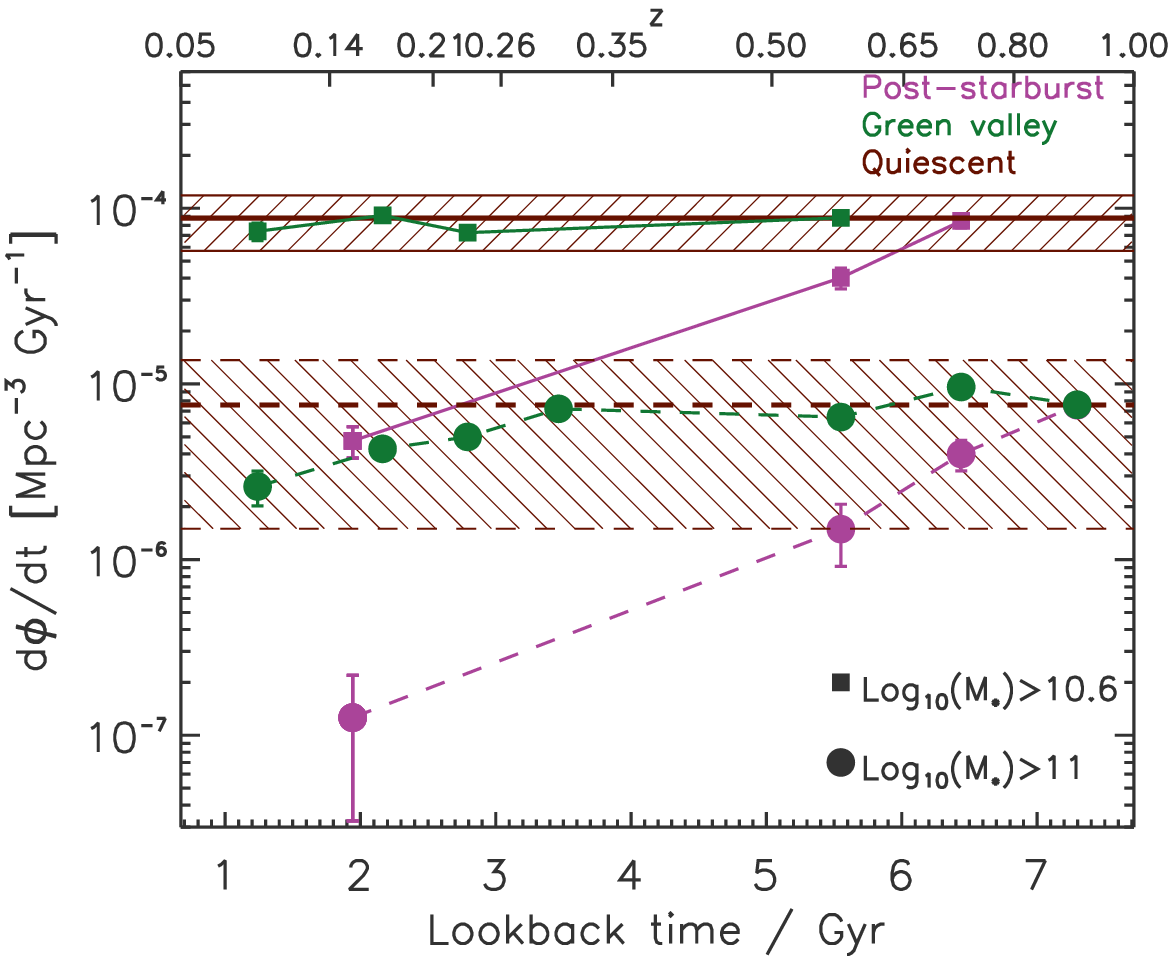}
\end{center}
\caption{The rate of galaxies passing through the PSB (purple) and green-valley (green) phase, and the growth rate of the quiescent population, as a function of redshift and mass. We assume a nominal transition time-scale of 2.6\,Gyr and 6.6\,Gyr for intermediate and high mass green-valley galaxies, respectively (see Section~\ref{sec:green}), and 0.5\,Gyr and 2.0\,Gyr for intermediate and high mass PSBs, respectively (see Section~\ref{sec:PSB}). Transition time-scales are estimated by requiring that $d\phi/dt_{\mathrm{Green valley}} \leq d\phi/dt_{\mathrm{Quiescent}}$ and $d\phi/dt_{\mathrm{PSB}} \leq d\phi/dt_{\mathrm{Quiescent}}$ in the highest redshift bin.
Results for intermediate mass ($\mstar>10^{10.6}$\msun) galaxies are shown as small squares connected by solid lines; high mass ($\mstar>10^{11}$\msun) galaxies are shown as large circles connected by dashed lines. The errors for the PSB and green valley populations include Poisson, cosmic variance contributions and uncertainties propagated from the uncertainty on the stellar masses.
The growth rate of the quiescent population is shown as a thick red line. The hatched area represents the uncertainty on the growth rate of the quiescent population, which is derived from the uncertainty on the linear fit to the number densities. The growth rate of star-forming galaxies of all masses is negative and is not shown.}
\label{fig:numbergrowth}
\end{figure}

We showed in Section~\ref{sec:NumberDensities} that the number density of the quiescent population is consistent with slow evolution at $0.05<z<1$ for \mstar$>10^{10.6}$\msun, and almost flat evolution for \mstar$>10^{11}$\msun.
We estimate the rate at which galaxies are entering the quiescent population ($d\phi/dt$) by fitting a straight line to the number density as a function of time to each mass bin of the quiescent population. The quantity $d\phi/dt$ is shown for intermediate ($\mstar>10^{10.6}$\msun) and high mass ($\mstar>10^{11}$\msun) quiescent galaxies in Figure~\ref{fig:numbergrowth} as the solid red and dashed lines, respectively. The hashed areas represents the upper and lower limits on the growth rate of the quiescent population, derived from the uncertainty on the linear fit to the quiescent population number densities as a function of time. While it is possible that the growth rate is not linear with time, the current data do not allow any higher order terms to be fit.
In the following discussion we neglect the effect of dry mergers which would cause quiescent galaxies to move within mass bins, as at these high masses the rate of dry mergers with close to equal mass ratios is expected to be small. While high mass quiescent galaxies may merge with low mass companions, they are not thought to merge with each other at low redshift to sufficiently affect their number density.

For the intermediate mass ($\mstar>10^{10.6}$\msun) quiescent galaxies we find a number density evolution with time which is growing at a rate of $d\phi/dt = 8.8\pm2.5  \times10^{-5}$Mpc$^{-3}$Gyr$^{-1}$.
For the high mass ($\mstar>10^{11}$\msun) quiescent galaxies we find a growth rate of $d\phi/dt = 7.6\pm6.7 \times10^{-6}$Mpc$^{-3}$Gyr$^{-1}$.
We note that aperture effects could cause us to classify galaxies differently depending on their redshift, particularly in the lowest redshift bin. We test the robustness of our results to redshift effects by repeating the growth rate calculation without the lowest redshift point and measure a slightly slower growth rate in both mass bins. However, the values are consistent within the $1\sigma$ uncertainties and our conclusions remain unchanged.

\subsection{Green valley galaxies}
\label{sec:green}
To put an upper limit on the rate at which green valley galaxies could be passing into the quiescent population ($d\phi/dt$), we divide the number densities in the intermediate mass bin by a nominal transition time-scale such that $d\phi/dt_{\mathrm{Green valley}} \leq d\phi/dt_{\mathrm{Quiescent}}$. This corresponds to a lower limit on the transition time-scale. To do this we make the assumptions that (1) galaxies cannot be transitioning faster than the growth rate of the quiescent population, and (2) that PSBs do not contribute to the growth of the quiescent population. In reality both green valley and PSB populations may be transitioning into the quiescent population, which would then require longer transition time-scales than given here. We discuss this further below.
Green valley galaxies with $\mstar>10^{10.6}$\msun\ are transitioning at a rate of $8\times10^{-5}$Mpc$^{-3}$Gyr$^{-1}$, for a transition time-scale of 2.6\,Gyr. Because the number density of intermediate mass green-valley galaxies remains constant with cosmic time, this transition time-scale refers to the rate at which galaxies pass through the boundaries we have defined for the green valley, assuming that every galaxy in this region is transitioning. If the transition time-scale does not change with redshift, the observed flat number density leads to a flat transition rate, which is consistent with a linear growth rate for the quiescent population. Similar conclusions of a relatively unchanging green valley population were found by \citet{Salim12}, \citet{Fang12}, and \citet{Salim15} who studied the UV morphologies and star-formation histories of green valley galaxies.

The estimated transition time-scale of $2.6^{+1.4}_{-0.7}$\,Gyr (accounting for the uncertainty on the growth rate of the quiescent population) is entirely reasonable, and is similar to time-scales found in the literature. Accounting for uncertainties, the transition time-scale for green valley galaxies cannot be $>4$\,Gyr otherwise $d\phi/dt_{\mathrm{Green valley}}$ will exceed $d\phi/dt_{\mathrm{Quiescent}}$. Using cosmological hydrodynamical simulations with radiation transfer post-processing, \citet{Trayford16} found using broad-band colours that most simulated galaxies spend $<2$\,Gyr in the green valley, independent of galaxy mass. \citet{Martin07} used spectral indices to estimate quenching times in local green valley galaxies, finding a time-scale of 50\,Myr to 6\,Gyr, with more than 50\% of quenching occurring within 2\,Gyr.
Using broad-band colours, \citet{Smethurst15} found a continuum of transition time-scales, but that most galaxies spend $1-2$\,Gyr in the green valley. Differences in transition time-scales may be due to the differences in selection methods of green valley galaxies (e.g. optical or UV colour-magnitude, spectroscopy).

In Figure~\ref{fig:numbergrowth} we show $d\phi/dt$ for high mass ($\mstar>10^{11}$\msun) green-valley galaxies, assuming a transition time-scale of $6.6^{+26.6}_{-2.9}$\,Gyr. Given the large errors on the growth of the high mass quiescent population, the drop in number density of the high mass green-valley galaxies remains consistent with a transition time-scale that does not evolve with redshift.
If we assumed a shorter time-scale of 2.6\,Gyr as used for the intermediate mass galaxies, the rate at which high mass galaxies pass through the green valley would be formally inconsistent with the growth rate of the quiescent population, after accounting for uncertainties at the $1\sigma$ level.
Our findings hint at an increasing transition time-scale with increasing mass, and may explain our slightly longer time-scale compared to other published values given our high mass limit ($\mstar>10^{10.6}$\msun).

Our results show that the growth of the quiescent population at these masses and redshifts can be entirely explained by galaxies transitioning slowly through the green valley. The presence of another, faster quenching channel would require a longer transition time for green-valley galaxies, or alternatively that only a fraction of the galaxies are actually transitioning.

\subsection{Post-starburst galaxies}
\label{sec:PSB}
We can put an upper limit on the rate at which galaxies could be passing through the PSB phase and into the quiescent population in the same way as for the green-valley galaxies. To reconcile the rate of intermediate mass ($\mstar>10^{10.6}$\msun) galaxies transitioning through the PSB phase at $z=0.7$ with the growth rate of the quiescent population, we  find a transition time-scale of 0.5\,Gyr. As above, this assumes that the green valley galaxies do not contribute to the growth of the quiescent population.

A transition time-scale of $0.5^{+0.3}_{-0.1}$\,Gyr (accounting for the uncertainty on the growth rate of the quiescent population) is a reasonable estimate for PSB galaxies and is similar to the time-scales found in hydrodynamical merger simulations by \citet{Wild09}, where the simulations are observed with the same spectral indices as used to identify PSB galaxies in this paper. Visibility time-scales of a few hundred Myrs for PSB features were also found in similar merger simulations by \citet{Snyder11}. Both papers found that the time-scales depend sensitively on gas fractions, orbital dynamics and progenitor types. The rapid decline in number density of post-starburst galaxies means that they must have significantly shorter visibility time-scales at low redshift if they are to contribute significantly to the growth of the quiescent population. While the simulations suggest that a visibility time-scale a factor of 2 shorter may be reasonable, this does not come close to the factor of 18 decrease in number density for intermediate mass galaxies. Equally, while it is possible that the rate of growth of the quiescent population slows between $z=1$ and $z=0.05$, and this is not captured by our linear fit, the change in number density of quiescent galaxies does not seem to indicate such a significant change. Aperture bias may cause us to select fewer transition galaxies at low redshift than at high redshift, but our tests in Section~\ref{sec:Aperture_bias} and Appendix~\ref{sec:colourmag} show that this is unlikely to cause such a strong evolution in number density as we observe. We can therefore conclude that while the fast-quenching post-starburst channel may contribute significantly to quiescent population growth of intermediate mass galaxies at $z=0.7$, it appears to be insignificant by $z\sim0$.

To reconcile the rate of high mass ($\mstar>10^{11}$\msun) galaxies transitioning through the PSB phase at $z=0.7$ with the growth rate of the quiescent population, we find a transition time-scale of $2.0^{+8.0}_{-0.9}$\,Gyr. This seems marginally inconsistent with the maximum possible visibility time-scale for post-starburst galaxies of $\sim$1\,Gyr (the main sequence lifetime of A-stars). Decreasing the time-scale to a more reasonable 0.5\,Gyr would give a transition rate that is incompatible with the observed growth of the quiescent population, especially when additional growth via the green-valley is included.
This indicates that at high masses some PSB galaxies may not be transitioning into the quiescent population. Further processes may be needed to fully quench these high-mass transition galaxies, i.e. they will subsequently return to the green valley or star-forming population. Alternatively, PSB galaxies may have rejuvenated from the quiescent population rather than transitioning from the star-forming population. The former scenario fits well with the findings that PSBs still have substantial gas and dust contents \citep{Zwaan13, French15, Rowlands15, Alatalo16b}, and often have disky morphologies \citep{Pawlik16}, indicating that they may not be fully quenched.

\subsection{Are galaxies quenching?}
The growth in number density of quiescent galaxies shows that quenching is occurring at $0.05<z<1$. The data is consistent with a quenching rate that is constant with cosmic time, and when combined with the observed number density of green valley galaxies, fits with a scenario in which the predominant quenching channel is the slow transitioning of green valley galaxies into the quiescent population over a time-scale of $\sim$2.6\,Gyr for galaxies with $\mstar>10^{10.6}$\msun, increasing to $\sim$6.6\,Gyr for galaxies with $\mstar>10^{11}$\msun. The existence of a significant number of PSB galaxies at $z\sim0.7$ draws this conclusion into question however. If both green valley and PSB galaxies contribute to the growth of the quiescent population at $z\sim0.7$ then the transition time-scales of both populations must be longer than the values estimated above. The maximum possible  time-scale for post-starburst galaxies is $\sim$1\,Gyr (the main sequence lifetime of A-stars). Therefore, at $z\sim0.7$ the growth of the intermediate mass quiescent population could conceivably be composed of, for example, an equal fraction of PSBs with a transition time of 1\,Gyr and green-valley galaxies with a transition time of $\sim$5.2\,Gyr. Although transition time-scales are generally expected to be shorter at higher redshift \citep{Goncalves12, Tinker13, Balogh16}, better data would be required to determine the exact rate at which the quiescent population is growing with redshift in order to rule out this scenario.

However, at high masses ($\mstar>10^{11}$\msun) there is clear tension between the number density of transition galaxies and small growth rate of the quiescent population. The transition times for the high mass galaxies are already long (2\,Gyr for PSBs and 6.6\,Gyr for green valley galaxies). If both green valley and PSB galaxies contribute to the growth of the quiescent population at $z\sim0.7$ then the transition time-scales of both populations will be unphysically long.
The large number of high mass transition galaxies compared to the slow growth of the quiescent population could be resolved if either, (i) the uncertainties in the mass growth of the quiescent population are underestimated, (ii) galaxies do not follow the linear evolutionary path of star forming, to quenching, to quiescent. The first scenario could be due to underestimation of cosmic variance, aperture bias, or systematics in the stellar masses due to IMF variation with redshift or galaxy mass. The first scenario will likely only be solved with larger spectroscopic surveys. Even though aperture bias may affect some classifications, the difference in the number densities between transition and quiescent galaxies is so large, and the evolution in the number densities of the green valley and PSB galaxies is so strong, that some misclassified galaxies are unlikely to affect our conclusions. Regarding the second scenario, several authors have suggested that PSB and green valley galaxies may have been rejuvenated and have temporarily come out of the quiescent population \citep{CorteseHughes09, Fang12, Dressler13}, and up to 60\% of local early-type galaxies have a cold ISM \citep[e.g.][]{Oosterloo10, Young11, Serra12, Rowlands12, Smith12, Agius13} which should allow them to rejuvenate given some trigger event. The broad-band colours of these gas-rich early-type galaxies are consistent with a rejuvenation scenario \citep{Young14}, where gas has been accreted recently via minor mergers \citep{Davis11}. Using cosmological hydrodynamical simulations with radiation transfer post-processing, \citet{Trayford16} found that 10\% of simulated galaxies are classified as rejuvenated as they show blue broad-band colours but were red in the past, although only 1.6\% of simulated galaxies rapidly change colour from red to blue over a $<2$\,Gyr time period.  However, \citet{Furlong15} showed that the passive fraction is too low in EAGLE at log($\mstar>10.5)$ by $\sim15$\% which may be a result of too much rejuvenation in their simulations. Direct comparisons of observations with simulations via the forward modelling of the simulations as mock datasets may help to unpick these related problems.
A temporary departure of galaxies from the quiescent population into the PSB or green-valley phase, possibly as a result of minor merger driven star formation, would relieve the tension between the number of candidate transition galaxies and the slow growth of the quiescent population at high masses. At intermediate masses, rejuvenation may happen, but is not visible compared to the number of truly quenching galaxies.

Alternatively, given the presence of a large cold ISM in PSBs \citep{Zwaan13, Rowlands15, French15, Alatalo16a}, high mass galaxies may originate from, and return to, the star forming population or green valley after a starburst. This was suggested by \citet{Dressler13} as the starburst galaxy population far outnumbers the passive galaxies in field and group environments. Overall, our findings for the highest mass ($\mstar>10^{11}$\msun) galaxies are in agreement with \citet{Dressler13} that the slow change in the numbers of quiescent galaxies since $z\sim1$ indicates that either not many galaxies in the PSB phase or green valley finally join the quiescent population, or that quenching takes multiple events happening over a long time.

Ultimately, since the growth rate of the quiescent population at $\mstar>10^{10.6}$\msun\ is slow, there is not a lot of room for the complete quenching of all massive galaxies since $z=1$, over half the age of the Universe. This means that although most galaxies are reducing their star formation rates with time, very few of them completely halt their star formation to become quiescent. At the highest masses ($\mstar>10^{11}$\msun), both rapid and slower quenching processes (e.g. strangulation and starvation) may be less effective. This implies that some fraction of populations commonly thought to be moving from star forming to quiescent, such as green valley and PSBs, may not be transitioning at all.

\section{Summary}
By exploiting the highly complete, wide-area GAMA and VIPERS spectroscopic surveys, we have studied the number densities of quiescent, PSB and green-valley galaxies. This has allowed us to explore the rate at which galaxies are quenching at $0.05<z<1$ as a function of mass, and the contribution of different transition galaxy populations to the build up of the quiescent population. Our main conclusions are summarised as follows:

\begin{itemize}

\item Over the last 8 billion years the quiescent population grows in number and mass density more quickly for intermediate mass ($\mstar>10^{10.6}\msun$) galaxies compared to high mass galaxies ($\mstar>10^{11}\msun$).

\item The number densities of spectroscopically classified green valley galaxies from $0<z<1$ is flat for intermediate mass and declining for high masses. We find the transition time-scale of intermediate (high) mass green valley galaxies is 2.6\,Gyr (6.6\,Gyr) at $0.05<z<1$ if green valley galaxies contribute 100\% to the growth of the quiescent population.

\item The number densities of PSB galaxies from $0<z<1$ is declining for both intermediate and high mass populations. The high mass PSB population shows a steeper decline in number density than at intermediate masses. We find that the transition time-scale of intermediate (high) mass PSBs is 0.5\,Gyr (2.0\,Gyr) at $z\sim0.7$ if PSB galaxies contribute 100\% to the growth of the quiescent population. The rapid decline in number density of PSBs with decreasing redshift means that the fast quenching channel must be insignificant by $z\sim0$.

\item If high redshift, intermediate mass PSB galaxies are visible for a slightly longer maximum time-scale of $\sim1$\,Gyr, corresponding to the main sequence lifetime of A stars, this allows quenching via both a slow and fast route to contribute to the growth of the intermediate mass quiescent population at $z\sim0.7$.

\item Both the green valley and PSB mass functions show that high mass galaxies transitioned at earlier cosmic times. The PSB mass function shows stronger redshift evolution than that of the green valley galaxy mass function.

\item The number of high mass ($\mstar>10^{11}\msun$) green valley and PSB galaxies is in tension with the observed slow growth of the quiescent population. This indicates that at high masses, some PSB or green valley galaxies are not transitioning from the star forming into the quiescent population. The mechanisms which cause a complete shut down in star formation may be rare or ineffective at $z<1$. Quiescent galaxies may undergo rejuvenation events which temporarily cause a galaxy to be observed in the green valley or PSB phase. Alternatively, the tension could be eased if the uncertainties on the number densities due to cosmic variance or stellar masses are underestimated.

\end{itemize}

This study has put upper limits on the rate at which galaxies are quenching, and by how much the slow and fast-quenching channels may be contributing at masses $\mstar>10^{10.6}\msun$ and redshifts in the range $0.05<z<1$. Larger spectroscopic surveys are needed to ascertain which quenching processes are acting in different environment and mass regimes. This issue will be addressed by the Taipan survey \citep{daCunha17}, which will measure fibre spectra for 1 million galaxies at $z<0.3$. Detailed information about the SFHs of candidate transition galaxies from high SNR spectra would confirm the fraction of the progenitors that are truly coming from the star-forming population, rather than rejuvenating. The processes which transform galaxies may leave different imprints on the motion and spatial distribution of stars and gas. Integral field spectroscopy from surveys such as MaNGA and SAMI could also help us to determine which processes cause quenching as a function of mass and environment.

\section*{acknowledgements}
We thank the referee for insightful and detailed comments which improved the paper.
We thank Alice Mortlock, Omar Almaini, Tim Davis and Alan Dressler for useful discussions. V.~W. and K.~R. acknowledge support from the
European Research Council Starting Grant SEDmorph (P.I. V.~Wild).
GAMA is a joint European-Australasian
project based around a spectroscopic campaign using the Anglo-Australian
Telescope. The GAMA input catalogue is based on data taken from the Sloan
Digital Sky Survey and the UKIRT Infrared Deep Sky Survey. Complementary imaging
of the GAMA regions is being obtained by a number of independent survey programs
including \emph{GALEX} MIS, VST KIDS, VISTA VIKING, \emph{WISE},
\emph{Herschel}-ATLAS, GMRT and ASKAP providing UV to radio coverage. GAMA is
funded by the STFC (UK), the ARC (Australia), the AAO, and the participating
institutions. The GAMA website is http://www.gama-survey.org/.
Based on observations made with ESO Telescopes at the La Silla Paranal
Observatory under programme ID 179.A-2004.
This paper uses data from the VIMOS Public Extragalactic Redshift Survey (VIPERS). VIPERS has been performed using the ESO Very Large Telescope, under the "Large Programme" 182.A-0886. The participating institutions and funding agencies are listed at http://vipers.inaf.it
This research has made use of NASA's Astrophysics Data System Bibliographic Services.

\bibliographystyle{mnras}
\bibliography{refs_all}

\begin{appendix}

\section{Removal of contaminants from the PSB population}
\label{sec:Unclean_PC12}
\begin{figure}
\includegraphics[width=0.48\textwidth]{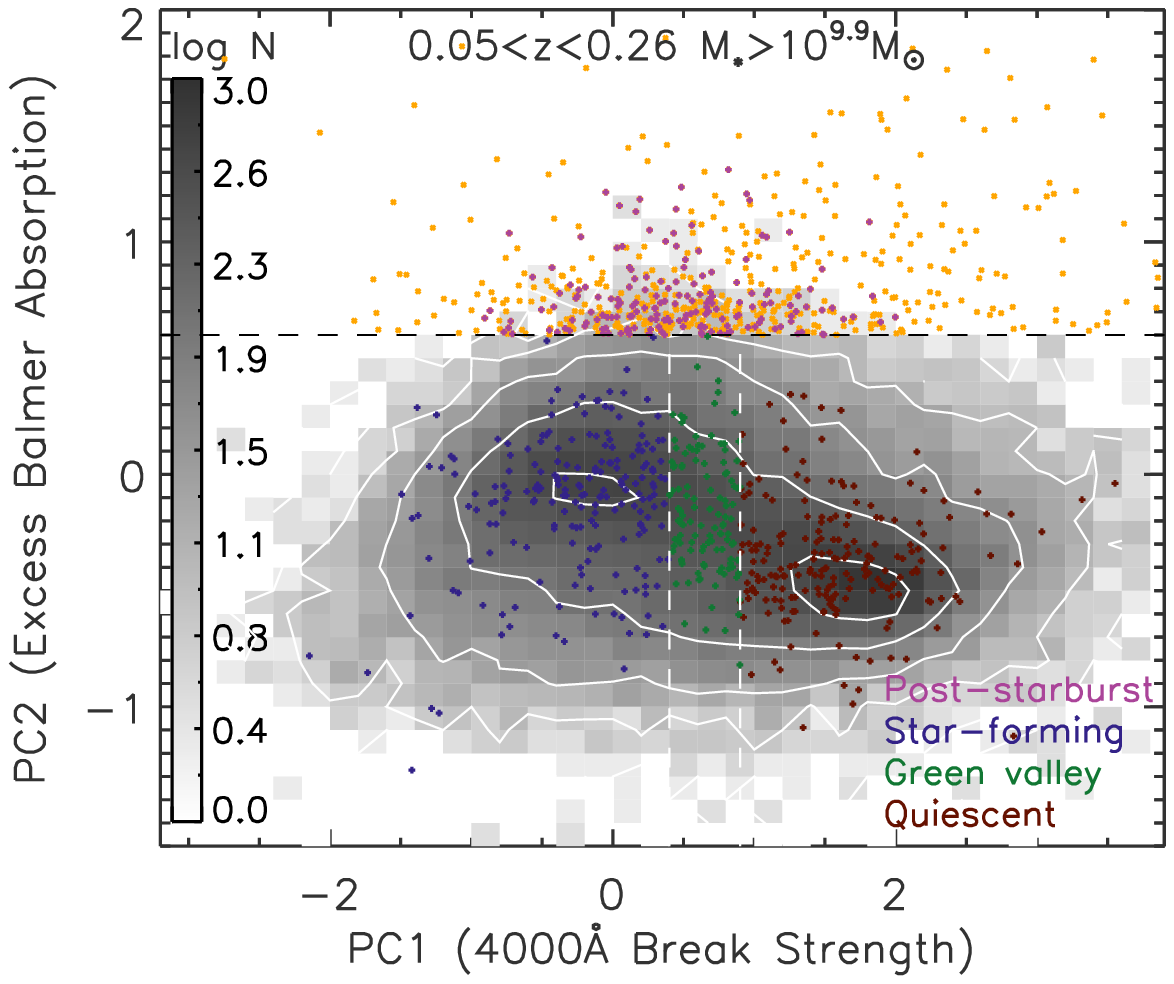}
\caption{The distribution of the 4000\AA\ break strength (PC1) and excess Balmer absorption (PC2) as measured using a Principal
Component Analysis of the 4000\AA\ spectral region of the GAMA galaxies in the range $0.05<z<0.26$. The grey-scale indicates the logarithmic number of objects. The coloured dots are random samples of galaxies which occupy each spectral class delineated by dashed lines: quiescent (red), star forming (blue), green valley (green) and clean PSB (purple), rejected PSB (orange). Contours show 10, 30, 50, 70 and 90\% of the maximum number of galaxies in the sample.}
\label{fig:GAMA_pc12_unclean}
\end{figure}

In Figure~\ref{fig:GAMA_pc12_unclean} we show the PCA selection of candidate PSBs (PC2$>0.6$) with SNR$>6.5$ in the redshift range $0.05<z<0.26$. Orange points show spectra which had problems with unmasked noise spikes, or exhibited an extreme fall off in flux to the blue (this could be due to poor tracing of the flux on the CCD when the SNR is low). Furthermore, some spectra in the PSB region were removed if we could not positively identify a Balmer series. The contaminant sources mostly occupy regions above the quiescent population and to the upper left of the star forming galaxies, with unphysical principal component amplitudes. Around $\sim 2/3$ of candidate PSB galaxies were removed, leaving a clean PSB sample (purple points), which lie in the expected PSB region above the star forming and green valley populations.

\section{Broad-band colours of spectrally classified galaxies}
\label{sec:colourmag}
\begin{figure*}
\begin{center}
\includegraphics[width=0.99\textwidth]{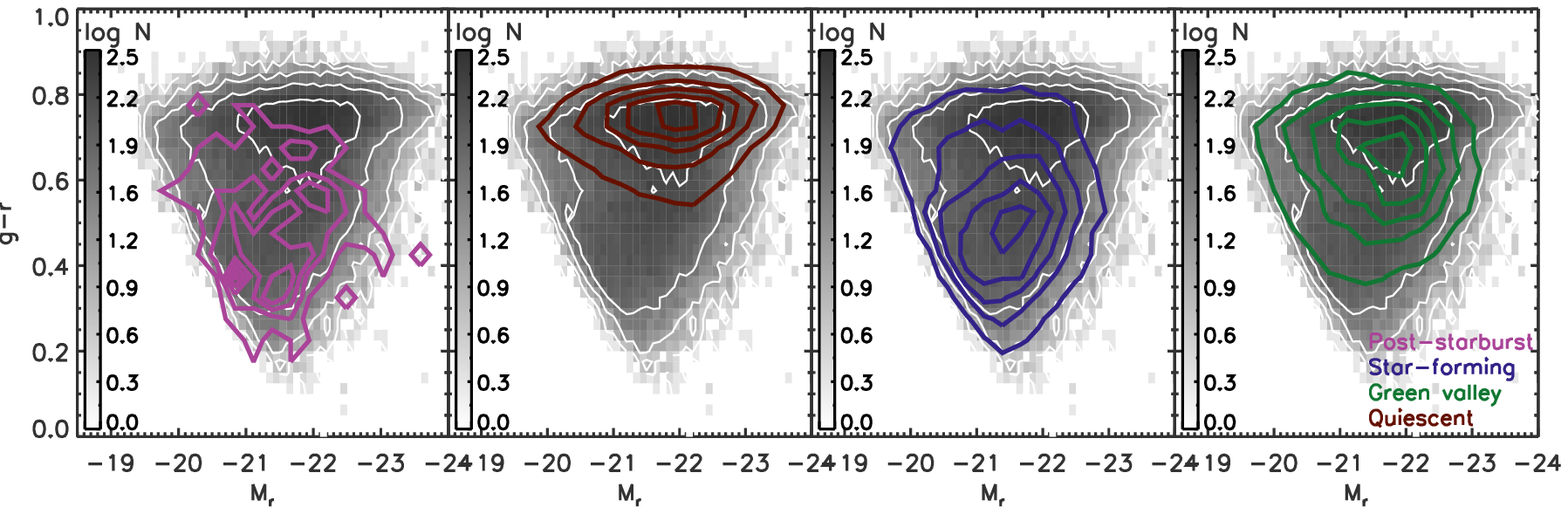}
\end{center}
\caption{Rest-frame broad-band colours as a function of absolute $r$ band magnitude of spectrally classified galaxies in the GAMA survey. The grey-scale indicates the logarithmic number of objects in the parent sample with SNR$>6.5$ per 6\AA\ pixel. Coloured contours show the different spectral classes in each panel, from left to right these are: post-starburst, quiescent, star-forming and green valley galaxies. Contours show 10, 30, 50, 70 and 90\% of the maximum number of galaxies in each sample.}
\label{fig:colour_mag}
\end{figure*}

In Figure~\ref{fig:colour_mag} we show the $g-r$ colours of galaxies in each spectral class. The broad-band colours have been K-corrected to rest-frame wavelengths using the best-fitting model SED from the \citet{BC03} library and have been corrected for Galactic extinction.
The broad-band colours of post-starburst galaxies are mostly in the $g-r$ blue cloud and green valley, with a minority of galaxies on the red sequence. The spectroscopically classified quiescent galaxies mostly reside in the red sequence, but there is some scatter towards the green valley. The spectroscopically classified green valley galaxies are centred on the $g-r$ green valley, but there is also substantial scatter into the blue cloud and red sequence. The spectroscopically classified star-forming galaxies lie mostly in the $g-r$ blue cloud but there is substantial scatter into the red sequence. We note that the definition of green valley based on broad-band photometry is arbitrary. The disagreement between the spectroscopic and photometric classification for some galaxies may be because of the different effects of dust on broad-band photometry and the information from performing the PCA on the spectra, which span a relatively narrow wavelength range of 3750-4150\AA. Furthermore, aperture bias will cause some differences, as the broad-band colours are an average over the whole galaxy, but the fibre only covers 2--3\arcsec{} of the galaxy centre. This is discussed further in Section~\ref{sec:Aperture_bias}.

\begin{figure*}
\begin{center}
\includegraphics[width=0.99\textwidth]{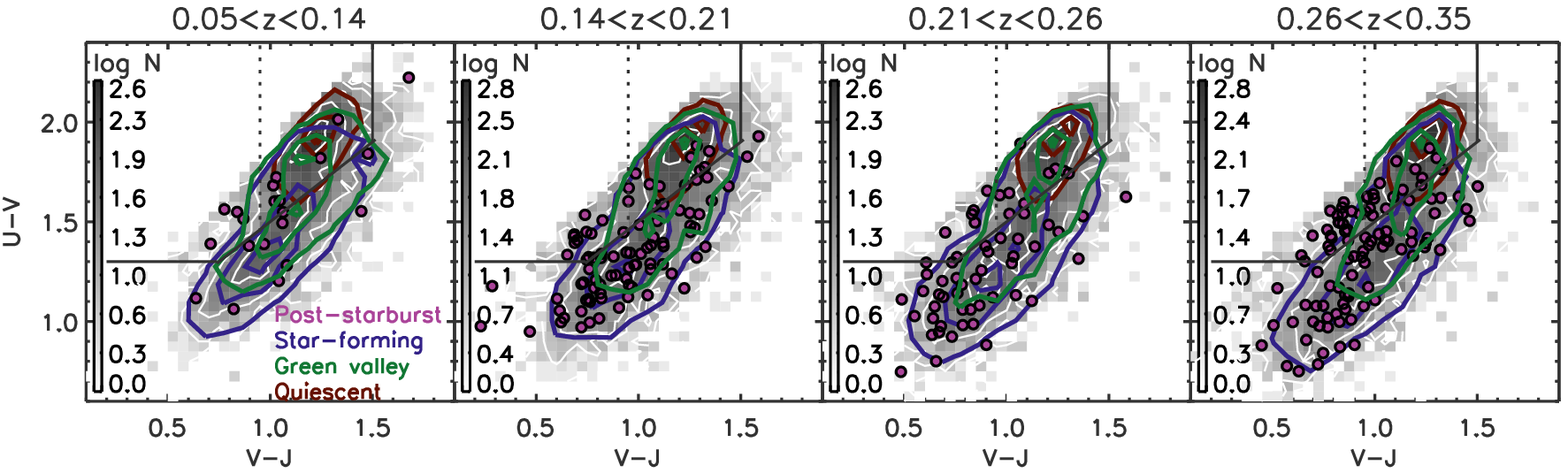}
\end{center}
\caption{Rest-frame broad-band colours of spectrally classified galaxies in the GAMA survey in four redshift bins. The grey-scale indicates the logarithmic number of objects in the parent sample with SNR$>6.5$ per 6\AA\ pixel. Coloured contours show the different spectral classes in each panel, from left to right these are: post-starburst, quiescent, star-forming and green valley galaxies. Contours show 10, 50, and 90\% of the maximum number of galaxies in each sample. Solid black lines are the commonly used separator between red-sequence, blue-sequence and dusty star-forming galaxies. The dotted
black line from \citet{Whitaker12a} separates post-starburst galaxies from red-sequence galaxies.}
\label{fig:UVJ}
\end{figure*}

Aperture bias is an issue which could affect our conclusions. To test this, in Figure~\ref{fig:UVJ} we show the U-V and V-J colours \citep[e.g.][]{Williams09} of the four spectral classes in four redshift bins. We compute the rest-frame $U$, $V$ and $J$ magnitudes by convolving the best-fit SED model (see Section~\ref{sec:stellar_masses}) with the Bessel $U$, Bessel $V$ \citep{Bessell_90} and UKIRT $J$ filters \citep{Tokunaga_02}.
Spectroscopically classified star-forming galaxies lie in the blue part of the UVJ diagram, whilst quiescent galaxies have red UVJ colours. Spectroscopically classified green valley galaxies have intermediate colours. The majority of PSBs overlap with the star-forming and green-valley populations. Galaxies in our sample have broad-band colours consistent with their spectroscopic classes, at all redshifts. If aperture bias was affecting our results we would see a shift in the locus of each spectral class from the broadband classification in a systematic way. We do not see any shift in the locus of any spectral class with redshift, therefore we see no evidence that aperture bias causes us to misclassify large numbers of galaxies of any spectral class.

\section{Comparison to literature mass functions}
\label{sec:MF_compare}

\begin{figure}
\begin{center}
\includegraphics[width=0.48\textwidth]{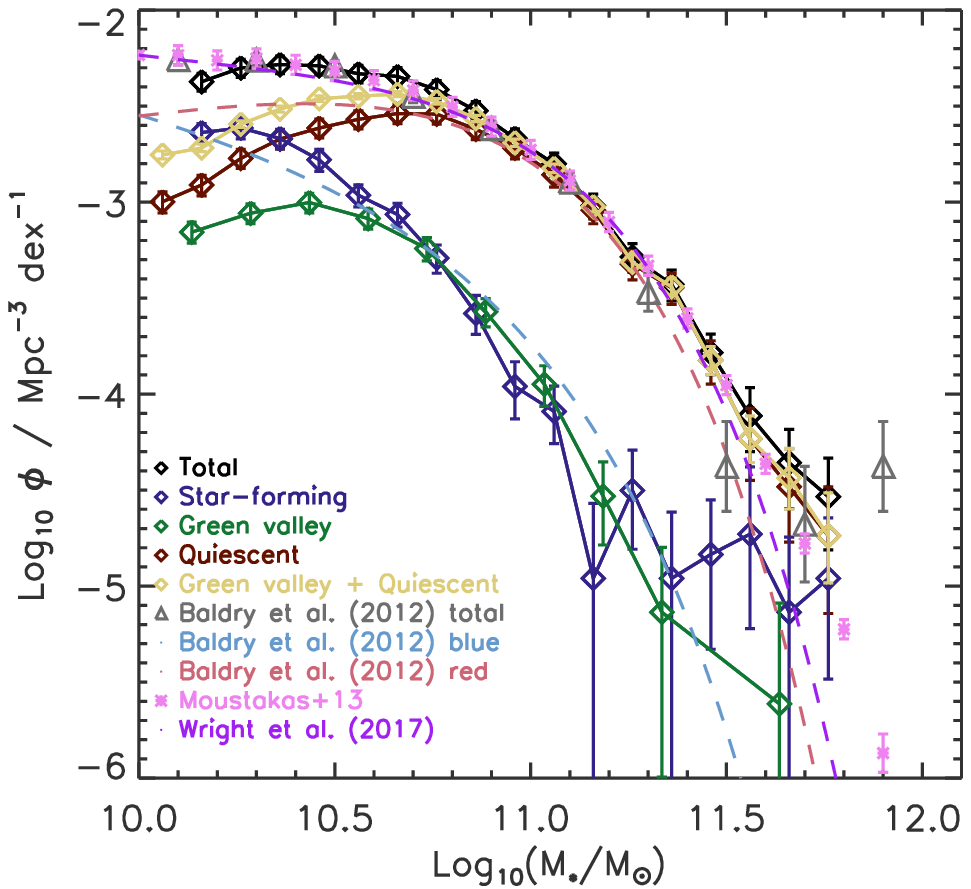}
\end{center}
\caption{The completeness corrected mass functions of the total, quiescent population, star-forming populations, green valley, and combined quiescent population and green valley populations for our lowest redshift bin ($0.05<z<0.14$) compared to those in the literature. Errors on mass functions from our study include Poisson, cosmic variance contributions and uncertainties propagated from the uncertainty on fitting the stellar masses.}
\label{fig:MF_compare_GAMA}
\end{figure}

In Figure~\ref{fig:MF_compare_GAMA} we compare our mass functions to those in the literature to check that our QSR completeness correction for spectra below our signal-to-noise threshold allows us to correctly recover the mass functions. Although the \citet{Baldry12} mass functions are derived from GAMA galaxies at $0<z<0.06$ and our sample spans $0.05<z<0.14$, there should be negligible evolution in this small redshift range. Our total and star-forming mass functions trace the \citet{Baldry12} mass functions well, which shows that our completeness corrections are correct. Furthermore, we find excellent agreement between our total mass function and that from \citet{Moustakas13} and \citet{Wright17}.

There is a slight deficit of low mass quiescent galaxies in our sample compared to \citet{Baldry12}. This may be due to the differences in sample selection, as \citet{Baldry12} separate star-forming and quiescent galaxies using the $u-r$ optical colour, and we use a cleaner spectroscopic selection that likely has less contamination by dusty objects. Furthermore, we separate quiescent from green valley galaxies, whereas \citet{Baldry12} do not make this discrimination, meaning that green valley galaxies will be mixed with the red and blue populations defined on broad-band colours. Adding the quiescent population and green valley mass functions together, they approximately reproduce the \citet{Baldry12} quiescent population mass function, with a slight deficit remaining as a small number of galaxies which are red in $u-r$ are classified spectroscopically as star-forming. The deficit of spectroscopic quiescent galaxies at the low mass end compared to \citet{Baldry12} suggests that broad-band optical colourselection tends to classify spectroscopic green-valley galaxies as red. Additionally, \citet{Baldry12} calculated stellar masses with $ugriz$ data only, whereas we derive more robust stellar mass measurements by using UV and NIR data in addition to $ugriz$ data, which may account for small differences in the mass functions. Note that \citet{Wright17} use updated GAMA photometry and we find excellent agreement with their mass function.

\begin{figure*}
\begin{center}
\includegraphics[width=0.99\textwidth]{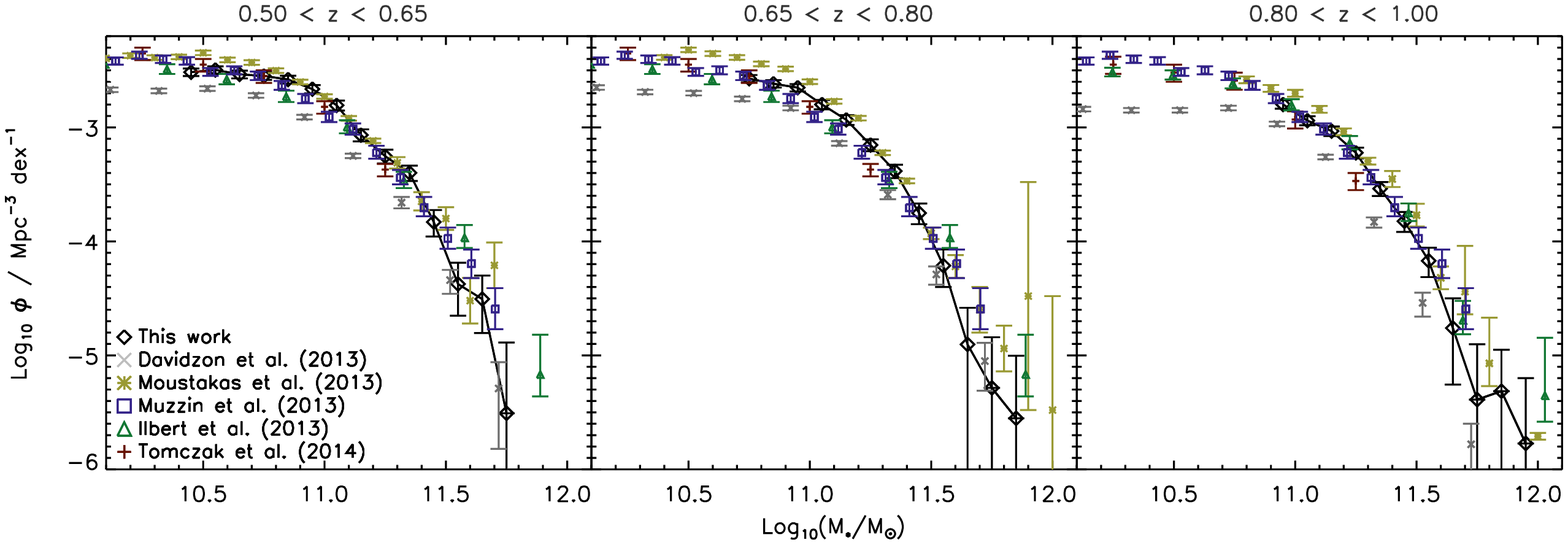}
\end{center}
\caption{The completeness corrected total mass functions in this work (black diamonds) compared to those in \citet{Davidzon13, Ilbert13, Moustakas13, Muzzin13, Tomczak14}. All points have been scaled to the same cosmology. Errors on the black points include Poisson, cosmic variance contributions and uncertainties propagated from the uncertainty on fitting the stellar masses.}
\label{fig:MF_compare_VIPERS}
\end{figure*}

In Figure~\ref{fig:MF_compare_VIPERS} we compare our VIPERS mass functions to those in \citet{Davidzon13}, \citet{Ilbert13}, \citet{Moustakas13}, \citet{Muzzin13}, and \citet{Tomczak14} using similar redshift ranges and cosmology. In all redshift bins we see good agreement with most literature studies. Our mass function has slightly different number densities in the $0.65<z<0.8$ bin, which may be caused by mismatches in the redshift ranges of literature studies, by different  stellar population models and/or fitting methods, or dust laws used by each study. Note that we use the same redshift ranges as \citet{Moustakas13} and we see an excellent agreement with this study in all redshift bins. There is no trend of worsening discrepancy with most literature mass functions with increasing redshift. However, after scaling our masses to the same cosmology, we find that the mass functions from \citet{Davidzon13} show an increasing discrepancy with redshift compared to our total mass functions. Whilst we do not use exactly the same redshift bins or stellar population models, the strong redshift trend suggests that this is unlikely to be the source of the difference between our mass functions. As the stellar masses from \citet{Davidzon13} are not publicly available we are  unable to dig deeper into this discrepancy.

\end{appendix}

\end{document}